\PassOptionsToPackage{hyphens}{url}
\documentclass[preprint,12pt,authoryear]{elsarticle}

\newcommand{\targetjournal}{Transportation Research Part A: Policy and Practice}
\journal{\targetjournal}

\usepackage{verbatim}

\usepackage[T1]{fontenc}
\usepackage[utf8]{inputenc}
\usepackage{newtxtext}
\IfFileExists{microtype.sty}{\usepackage{microtype}}{} 

\makeatletter
\ifnum\jtype=0
  \usepackage[a4paper, left=1.1in, right=1.1in, top=0.98in, bottom=0.98in]{geometry}
\fi
\makeatother

\makeatletter
\renewcommand\section{\@startsection{section}{1}{\z@}%
  {18\p@ \@plus 6\p@ \@minus 3\p@}{9\p@ \@plus 6\p@ \@minus 3\p@}%
  {\normalfont\large\bfseries\boldmath}}
\renewcommand\subsection{\@startsection{subsection}{2}{\z@}%
  {12\p@ \@plus 6\p@ \@minus 3\p@}{3\p@ \@plus 6\p@ \@minus 3\p@}%
  {\normalfont\normalsize\bfseries\boldmath}}
\renewcommand\subsubsection{\@startsection{subsubsection}{3}{\z@}%
  {12\p@ \@plus 6\p@ \@minus 3\p@}{\p@}%
  {\normalfont\normalsize\itshape}}
\makeatother
\usepackage{amsmath,amsthm}
\usepackage{newtxmath}
\usepackage{bm}               
\usepackage{mathtools}        

\usepackage{graphicx}
\graphicspath{{figures/}}
\usepackage{subcaption}       
\usepackage{tikz}             
\usetikzlibrary{arrows.meta,positioning,shapes.geometric,calc}
\usepackage[section]{placeins}        

\usepackage{booktabs}         
\usepackage{threeparttable}   
\usepackage{multirow}
\usepackage{array}
\usepackage{adjustbox}        
\usepackage{siunitx}          
\newcolumntype{L}[1]{>{\raggedright\arraybackslash}p{#1}}
\newcolumntype{C}[1]{>{\centering\arraybackslash}p{#1}}
\newcolumntype{R}[1]{>{\raggedleft\arraybackslash}p{#1}}

\usepackage{xltabular}        
\newcolumntype{Y}{>{\raggedright\arraybackslash}X}
\usepackage{etoolbox}
\makeatletter
\AtBeginEnvironment{highlights}{\ifdoubleblind\fi}
\AtBeginEnvironment{graphicalabstract}{\ifdoubleblind\fi}
\makeatother

\usepackage{enumitem}
\setlist{nosep}
\usepackage{xcolor}
\usepackage[mathlines]{lineno} 

\definecolor{elslink}{RGB}{73,148,204}
\usepackage{hyperref}         
\hypersetup{
  colorlinks = true,
  pdftitle   = {Manuscript title},
  pdfauthor  = {First Author; Second Author},
}
\AtBeginDocument{\hypersetup{
  citecolor  = elslink,       
  linkcolor  = elslink,       
  urlcolor   = elslink,       
}}
\makeatletter
\AtBeginDocument{%
  \def\hyper@natlinkbreak#1#2{#1}%
  \patchcmd{\@citex}{\relax\else\NAT@@close\fi\fi}%
    {\relax\else\ifHy@colorlinks{\color{\@citecolor}\NAT@@close}\else\NAT@@close\fi\fi\fi}%
    {}{\PackageWarning{packages.tex}{natbib closing-parenthesis patch failed}}%
}
\makeatother
\usepackage[capitalise,noabbrev,nameinlink]{cleveref}
\crefname{equation}{Eq.}{Eqs.}
\Crefname{equation}{Eq.}{Eqs.}
\creflabelformat{equation}{#2(#1)#3}
\crefrangelabelformat{equation}{#3(#1)#4--#5(#2)#6}

\makeatletter
\newcommand*{\appendixseccntformat}[1]{\csname the#1\endcsname\hspace{0.6em}}
\makeatother

\makeatletter
\apptocmd{\appendix}{%
  \let\@seccntformat\appendixseccntformat%
  \setcounter{figure}{0}%
  \setcounter{table}{0}%
  \setcounter{equation}{0}\renewcommand{\theequation}{A.\arabic{equation}}%
  \crefalias{section}{appendix}%
  \crefalias{subsection}{subappendix}%
  \crefalias{subsubsection}{subsubappendix}%
}{}{\PackageWarning{packages.tex}{could not patch \string\appendix}}
\makeatother

\crefformat{appendix}{#2#1#3}
\Crefformat{appendix}{#2#1#3}
\crefrangeformat{appendix}{#3#1#4--#5#2#6}
\Crefrangeformat{appendix}{#3#1#4--#5#2#6}
\crefmultiformat{appendix}{#2#1#3}{ and #2#1#3}{, #2#1#3}{ and #2#1#3}
\Crefmultiformat{appendix}{#2#1#3}{ and #2#1#3}{, #2#1#3}{ and #2#1#3}

\crefformat{subappendix}{#2#1#3}
\Crefformat{subappendix}{#2#1#3}
\crefrangeformat{subappendix}{#3#1#4--#5#2#6}
\Crefrangeformat{subappendix}{#3#1#4--#5#2#6}
\crefmultiformat{subappendix}{#2#1#3}{ and #2#1#3}{, #2#1#3}{ and #2#1#3}
\Crefmultiformat{subappendix}{#2#1#3}{ and #2#1#3}{, #2#1#3}{ and #2#1#3}

\crefformat{subsubappendix}{#2#1#3}
\Crefformat{subsubappendix}{#2#1#3}
\crefrangeformat{subsubappendix}{#3#1#4--#5#2#6}
\Crefrangeformat{subsubappendix}{#3#1#4--#5#2#6}
\crefmultiformat{subsubappendix}{#2#1#3}{ and #2#1#3}{, #2#1#3}{ and #2#1#3}
\Crefmultiformat{subsubappendix}{#2#1#3}{ and #2#1#3}{, #2#1#3}{ and #2#1#3}

\usepackage{float}
\floatstyle{ruled}                 
\newfloat{algorithm}{tbp}{loa}
\floatname{algorithm}{Algorithm}
\crefname{algorithm}{Algorithm}{Algorithms}
\Crefname{algorithm}{Algorithm}{Algorithms}

\newtheorem{proposition}{Proposition}   

\IfFileExists{dsfont.sty}{\usepackage{dsfont}}%
                         {}
\newcommand{\vect}[1]{\bm{#1}}              
\newcommand{\T}{^{\mathsf{T}}}              

\newcommand{\eps}{\varepsilon}

\DeclareSIUnit{\vehicle}{veh}
\DeclareSIUnit{\trip}{trip}
\DeclareSIUnit{\pax}{pax}

\DeclareMathOperator*{\argmin}{arg\,min}

\makeatletter
\def\@afterindentfalse{\let\if@afterindent\iffalse}
\let\@afterindenttrue\@afterindentfalse
\@afterindentfalse
\makeatother     

\begin{document}


\begin{frontmatter}

\title{AMXL-z: Modeling taste heterogeneity in large-scale GPS data with an application to public EV charging} 



\author[C2SMART,CUE]{Xiyuan (Ryan) Ren}
\ead{xr2006@nyu.edu}

\author[CUSP,Marron]{Callie Clark}

\author[NYU Shanghai]{Yunyi (Lauren) Wang}

\author[CUSP,TMI]{Takahiro Yabe}

\author[C2SMART,CUE]{Joseph Y. J. Chow\corref{cor1}}
\ead{joseph.chow@nyu.edu}

\cortext[cor1]{Corresponding author.}

\affiliation[C2SMART]{organization={C2SMART Center, Tandon School of Engineering, New York University},
  country={USA}}
  
\affiliation[CUE]{organization={Department of Civil, Urban, and Environmental Engineering, Tandon School of Engineering, New York University},
  country={USA}}

\affiliation[CUSP]{organization={Center for Urban Science + Progress (CUSP), Tandon School of Engineering, New York University},
  country={USA}}

\affiliation[Marron]{organization={Marron Institute for Urban Management, New York University},
  country={USA}}

\affiliation[NYU Shanghai]{organization={New York University Shanghai},
  country={China}}


\affiliation[TMI]{organization={Department of Technology Management and Innovation (TMI), Tandon School of Engineering, New York University},
  country={USA}}


\begin{abstract}
GPS data provide large samples with rich panel information on travel choices, but individual characteristics are often limited, aggregated, or noisy. We propose an Agent-based Mixed Logit model with observed characteristics (AMXL-z) to address this trade-off by decomposing agent-level tastes into a population baseline, systematic variation explained by observed characteristics, and residual heterogeneity learned empirically from panel data. An agent-level estimation framework further enables AMXL-z to scale efficiently to large datasets. Simulation experiments show that AMXL-z achieves competitive predictive performance and parameter recovery relative to the neural-network-based TasteNet, particularly when observed characteristics are noisy or panel choice histories are available. Compared with conventional mixed logit, AMXL-z maintains substantially more stable memory requirements as panel size increases. 
We demonstrate the method by modeling joint public EV charging-location and nearby activity choices using 8,930 charging sessions from 1,747 inferred EV drivers in the San Francisco Bay Area. Session-specific choice sets are constructed from activity chains using a space-time prism, and nearby activity utility enters charging-station choice through a nested structure. The estimated trade-off between detour time and charging price corresponds to 33.9 minutes per \$1/kWh reduction and an implied detour-time value of \$18.1/hour. AMXL-z further reveals substantial differences in the sources of taste heterogeneity. Observed driver and trip-context variables explain 62.5\% and 73.2\% of variation in access-distance and detour tastes, but only 4.1\% for price, 10.0\% for DC fast-charging availability, and 2.2\% for nearby activity opportunities. These results demonstrate how separating systematic and residual heterogeneity can uncover behavioral structure that would otherwise remain obscured and inform more targeted charging-infrastructure and behavioral policies.

\end{abstract}


\end{frontmatter}

\section{Introduction}
\label{sec:intro}
The transition toward electric mobility is placing greater emphasis on the deployment of public electric vehicle charging stations (PEVCSs). In the United States, federal and state initiatives have accelerated the expansion of charging infrastructure toward the national target of 500,000 EV chargers by 2030 \citep{fhwa2022,whitehouse2023}, with PEVCSs projected to serve at least 30\% of total charging demand \citep{pierce2023home}. As public charging networks expand, decisions about where and how charging facilities are deployed will increasingly influence not only charging accessibility and EV adoption, but also surrounding travel and activity patterns. 

Public charging differs fundamentally from conventional refueling because drivers remain at charging locations for substantially longer periods, often from 20 minutes to more than an hour even with fast charging \citep{dot2023charger}. This dwell time makes charging not just a travel-related task, but an activity itself. Recent studies have therefore begun to integrate charging with daily activity scheduling \citep{pan2020location,wang2025joint}, or measure causal links between charging infrastructure and nearby business activity \citep{zheng2024effects,de2025causal}. Yet it remains unclear whether and to what extent the utility of nearby activity opportunities enters drivers’ choices among feasible PEVCSs, and how this effect varies across drivers and trip characteristics.

Discrete choice models (DCMs) have been widely used to quantify drivers’ preferences for charging stations. Existing studies have identified the importance of charging speed, price, waiting time, accessibility, and nearby amenities \citep{huang2026preferences,yao2026modeling,potoglou2023public}. However, most studies rely on stated preference (SP) experiments, in which preferences are estimated under hypothetical choice scenarios \citep{sica2025electric,visaria2022user}. A small number of revealed-preference (RP) studies examine observed charging decisions, but generally separate charging location choice from activities undertaken during charging \citep{lee2020exploring,kim2017heterogeneous}. The joint charging-location–activity mechanism therefore remains largely unobserved.

The increasing availability of large-scale GPS data creates a distinct methodological challenge for modeling taste heterogeneity \citep{wu2025planning}. Such data provide large samples with rich panel information on travel choices and make it possible to construct realistic choice contexts from observed travel and activity sequences \citep{gonzalez2008,xu2018planning,fu2025activity,clark2026using}. At the same time, individual characteristics in passively collected mobility data are often limited, spatially aggregated, or noisy. As a result, only part of taste heterogeneity can be systematically associated with observed characteristics, while the remaining variation should be empirically learned from the data.

Existing DCMs address these two sources of heterogeneity only partially. Conventional models such as multinomial logit (MNL) and mixed logit (MXL) represent preferences through fixed coefficients or prespecified parametric distributions \citep{yao2026modeling,sica2025electric}. More flexible approaches have recently been developed, including nonparametric models such as agent-based mixed logit (AMXL) \citep{ren2022random,ren2025nonparametric} and individual parameter logit \citep{swait2023distribution}, as well as neural-network-based models such as TasteNet-logit \citep{han2022neural}, GNN-DCMs \citep{cheng2026graph}, and MDN-Logit \citep{li2026decomposing}. These approaches relax some of the restrictions imposed by conventional logit models, but a trade-off remains between explaining taste variation through observed characteristics and flexibly recovering the heterogeneity that those characteristics cannot explain. This trade-off is particularly important for large-scale GPS data, where rich panel choice histories coexist with incomplete or noisy individual information.

To address this challenge, we extend the AMXL model \citep{ren2022random} to Agent-based Mixed Logit with observed characteristics (AMXL-z). AMXL-z decomposes agent-level tastes into a population baseline, systematic variation explained by observed characteristics ($\vect{z}$), and residual heterogeneity learned empirically from panel data. This structure allows the model to exploit the rich panel information available in large-scale GPS data without requiring all taste heterogeneity to be explained by potentially incomplete or noisy observed characteristics. An agent-level estimation framework further allows the model to scale to large datasets. We evaluate AMXL-z through a simulation study against conventional MXL and the neural-network-based TasteNet-logit, focusing on predictive performance, recovery of taste heterogeneity, and computational efficiency. To facilitate future research, we upload the simulation experiments to a \href{https://github.com/xr2006/AMXL-z-simulation-study}{GitHub repository}.

We then demonstrate the proposed framework through public EV charging behavior in the San Francisco Bay Area. Using 8,930 charging sessions made by 1,747 inferred EV drivers identified by \cite{clark2026using} from large-scale mobile-phone GPS data, we construct session-specific charging choice sets from observed activity chains using a space-time prism approach. Candidate PEVCSs are determined by the preceding and subsequent activity locations, available time budget, and detour constraints, allowing the choice set to reflect the actual context faced by each driver. We further develop a nested choice framework in which the lower branch represents point-of-interest (POI) visitation as a function of the surrounding built environment, while its expected utility enters the upper-level charging-location choice together with travel and station attributes. This empirical application allows us to examine not only whether nearby activity opportunities affect charging-station choice, but also which components of charging and activity preferences can be systematically explained by observed driver and trip contexts and which remain as residual taste heterogeneity.

The remainder of the paper is organized as follows. \Cref{sec:lit} reviews studies on publich EV charging behavior and DCMs that estimate taste heterogeneity. \Cref{sec:methods} outlines the methodology, including choice set generation, model specification, and estimation algorithm. \Cref{sec:simulation} evaluates the proposed AMXL-z model against MNL, MXL, and TasteNet using simulated data. \Cref{sec:case} presents the model results for the San Francisco Bay Area. \cref{sec:conclusion} concludes with key takeaways and directions for future research.
\section{Literature Review}
\label{sec:lit}


\subsection{Public EV charging behavior}
\label{sec:lit-PEVCS}

Existing studies on public EV charging have examined several related choice dimensions. Some focus on choices across home, workplace, public charging site, or a mix of locations \citep{lee2020exploring,chakraborty2019demand}, while others examine choices of charging (or not) at public charging stations and charger types \citep{li2023uncertainty,ge2022charging}. A growing body of work compares choices of en-route and destination charging \citep{mehditabrizi2025route}, or studies charging under such specific trip contexts as commuting, long-distance travel, shopping, and recreation \citep{yang2025dominant,wang2025joint,wang2026revisiting}. Most studies rely on SP experiments with hypothetical charging scenarios \citep{potoglou2023public} in which respondents involve current and potential EV drivers \citep{gutjar2023configuration}. Discrete choice models (DCMs) have been a common framework for studying how drivers evaluate public charging alternatives, with applications including binary logit \citep{li2023uncertainty}, multinomial logit (MNL) \citep{gutjar2023configuration}, mixed logit (MXL) \citep{visaria2022user}, and latent class (LC) models \citep{ge2022charging}.

A consistent finding across this literature is that public charging choices are shaped by a broad set of factors. Station attributes such as charging speed, price, waiting time, charger availability, and number of charging points determine the competitiveness of a charging station \citep{visaria2022user,ge2022charging}. Nearby amenities also play an important role, particularly for longer charging sessions, with previous studies finding preferences for stations near shopping, dining, and other services \citep{gazmeh2026misalignment}. Vehicle conditions, including state of charge, remaining driving range, and battery capacity, further influence the need and urgency to charge \citep{chakraborty2019demand}. Temporal factors such as time of day, weekday versus weekend, and weather condition affect sensitivities to waiting time and charging prices \citep{gellrich2022spatial}. Activity schedules and trip chains further constrain when and where charging can occur, linking charging decisions to broader daily travel patterns \citep{wang2025joint}.  

Moreover, the effects of these attributes are heterogeneous across drivers and contexts. Previous studies have shown that taste parameters associated with charging attributes vary with demographic and socioeconomic characteristics, EV experience, vehicle ownership, and travel context \citep{potoglou2023public}. For example, more experienced EV drivers are more likely to use public charging \citep{wang2021electric}, while frequent public-charging users tend to prefer fast charging \citep{ten2020investigating}. In addition, differences in public charging choices can be explained by timing of charging and risk-taking attitudes \citep{wang2021electric,pan2019modeling,wen2016modeling}. Existing models have represented this heterogeneity in different ways. MNL models typically capture taste heterogeneity through interactions with individual characteristics \citep{gutjar2023configuration}, while MXL models introduce parametric distribution in taste parameters \citep{visaria2022user}. LC models instead identify a finite number of groups with distinct charging preferences \citep{ge2022charging}.

Despite this progress, the empirical basis for estimating charging preferences remains limited. Only a small number of studies have used RP data from observed charging decisions \citep{lee2020exploring,kim2017heterogeneous}. Large-scale GPS data have become increasingly available, but these datasets have often been used to describe aggregate usage patterns rather than to construct real choice contexts and available alternatives faced by individual EV drivers. Consequently, relatively little is known about taste heterogeneity revealed through large-scale mobility traces, particularly when preferences may vary jointly with travel context, time constraints, and surrounding activity opportunities. As EV adoption continues to grow, public charging infrastructure planning should continuously adapt to drivers’ evolving behavior, motivating more flexible choice modeling approaches that can learn heterogeneous charging preferences from emerging mobility datasets.

\subsection{Preference heterogeneity in choice models}
\label{sec:lit-DCM}

Preference heterogeneity in choice models refers to variation in the taste parameters that determine how decision makers value alternative attributes \citep{mcfadden2000mixed}. Consider the utility function
$ U_{nit} = \vect{x}_{nit}\T \vect{\beta}_{nt} + \eps_{nit}$
where $\vect{x}_{nit}$ denotes the attributes of alternative $i$, and
$\vect{\beta}_{nt}$ represents the taste parameters of individual $n$ at choice
occasion $t$. These parameters can be expressed as
$\vect{\beta}_{nt} = \bar{\vect{\beta}} + g(\vect{z}_{nt}) + \vect{\eta}_{nt}$
where $\vect{z}_{nt}$ contains observed individual characteristics and
choice-context variables, and $\vect{\eta}_{nt}$ captures unobserved preference
variation. 

Conventional DCMs (e.g., MNL, MXL, and LC) impose relatively strong
assumptions on both components. The effects of observed characteristics $\vect{z}_{nt}$ are typically introduced through prespecified interaction terms, requiring researchers to assume which variables interact with which taste parameters and often restricting these relationships to linear forms \citep{bowman2001activity}. Unobserved heterogeneity is commonly assumed to follow parametric distributions, such as Gaussian or lognormal distributions \citep{mcfadden2000mixed}, rather than learned directly from the data. Although MXL can recover individual-specific taste parameters through maximum a posteriori (MAP) estimation using panel choice data \citep{train2009}, it remains unclear to what extent this posterior updating can offset biases introduced by simplified interactions and parametric distributions. 

Semi-parametric and nonparametric approaches have been developed to relax the distributional assumptions imposed on unobserved taste heterogeneity. \cite{train2016mixed} proposed the logit-mixed logit (LML) model, in which the mixing distribution can be flexibly represented using splines, polynomials, and other functional forms. \cite{ren2022random} developed the agent-based mixed logit (AMXL) model to recover deterministic individual-specific parameters through a hybrid econometric and machine-learning procedure. \cite{swait2023distribution} introduced a nonparametric estimator that directly maximizes the sample likelihood to obtain individual-level preferences. These approaches allow the distribution of taste parameters to be learned more flexibly from observed choices without imposing strong assumptions on parametric distributions. However, the estimated heterogeneity generally combines the systematic variation, $g(\vect{z})$, and unobserved variation, $\vect{\eta}$, into an empirical distribution of $\vect{\beta}$. The two sources of heterogeneity are therefore not explicitly separated during estimation, and the relationship between observed characteristics and estimated tastes can only be examined ex post.

The application of machine learning in travel behavior analysis has accelerated a significant stream of studies on deep neural network (DNN) choice models \citep{wang2020deep,sifringer2020enhancing}. \cite{sifringer2020enhancing} proposed the learning multinomial logit (L-MNL) model which captures the mapping between $\vect{z}$ and utility functions through a 2D convolutional neural network. \cite{han2022neural} extended L-MNL and introduced TasteNet, which allows the neural network to learn the interactions between $\vect{z}$ and taste parameters. \cite{cheng2026graph} developed graph-neural-network-based DCMs to incorporate dependence across alternatives and showed that nested logit can be represented as a special case of their framework. \cite{li2026decomposing} developed MDN-Logit, which uses a mixture density network (MDN) to learn conditional distributions of taste parameters, instead of relying on point estimates. These approaches can effectively reduce the share of preference variation left in the residual component $\vect{\eta}$ by explaining more of it through individual characteristics, alternative dependence, and contextual variation. However, DNN-based models depend heavily on the richness of the observed $\vect{z}$ vector and can be sensitive to hyperparameter settings and random initialization, increasing computational cost and introducing additional estimation uncertainty. Moreover, repeated observations from the same individual are not always exploited as directly as in panel-based mixed logit models, which may limit how efficiently these models learn from longitudinal choice data.

The pros and cons of large-scale GPS data shape the requirements for modeling public charging behavior. On the one hand, these data provide large samples and repeated observations of the same individuals. Models designed for such data therefore need to scale efficiently with sample size, exploit panel structure, and keep memory use and computational cost manageable. On the other hand, privacy protection often limits the availability of individual characteristics in the observed $\vect{z}$ vector. Information such as age, income, and education may only be available at an aggregated spatial level and therefore contain non-negligible measurement error. This limits the extent to which $\vect{z}$ can explain taste variation and suggests that the remaining unobserved heterogeneity should not be assumed to follow a parametric distribution. Together, these properties motivate choice models that can learn empirical preference distributions while making efficient use of large-scale panel data and remaining robust to incomplete or noisy observed characteristics.

\section{Methodology}
\label{sec:methods}
\begin{figure}[htbp]
  \centering
  \includegraphics[width=\linewidth]{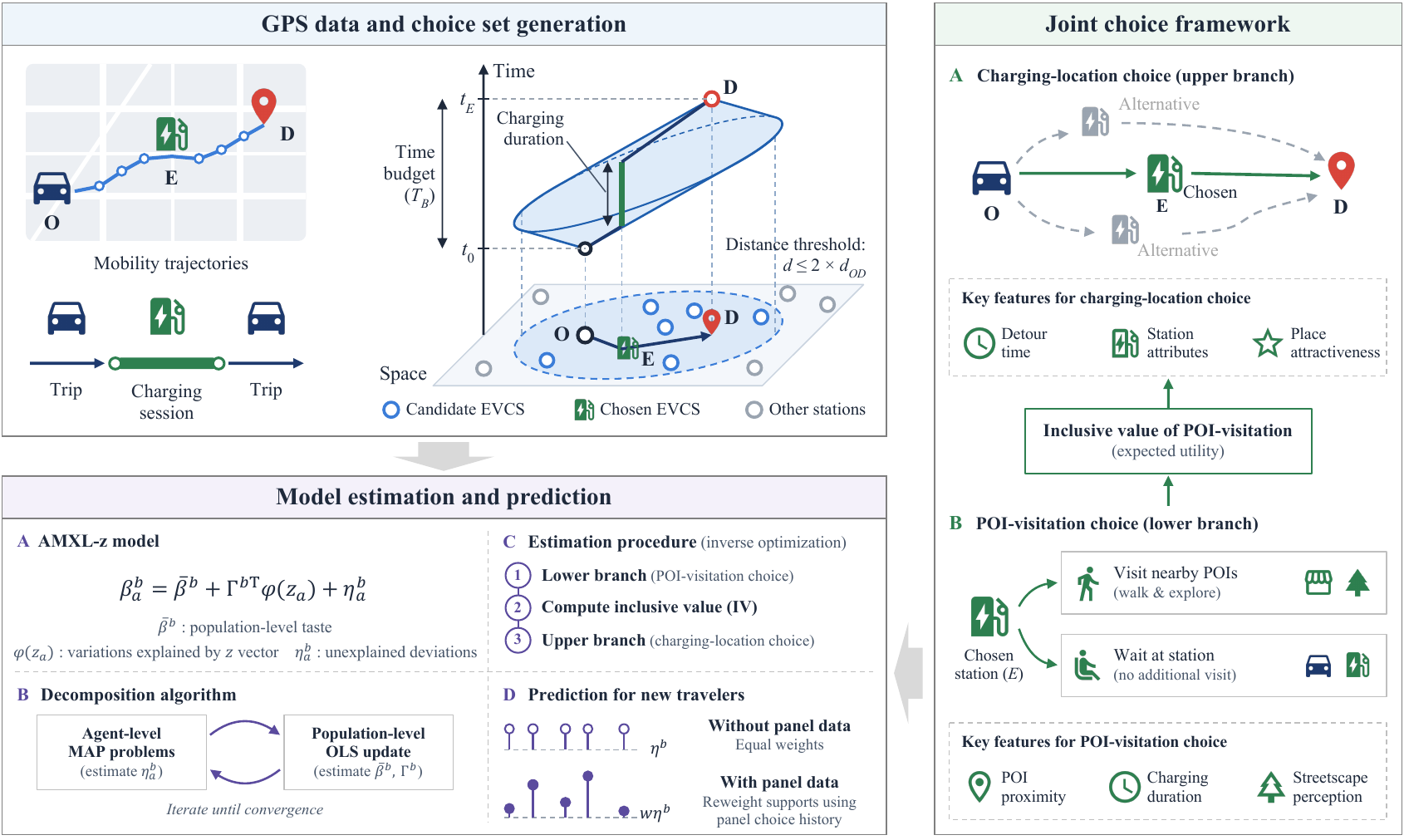}
  \caption{Overview of the proposed framework for modeling joint charging location and activity choice with AMXl-z using GPS data.}
  \label{fig:framework}
\end{figure}

The proposed methodology integrates three parts as shown in \cref{fig:framework}: (1) Choice set generation, (2) joint charging decision framework, and (3) estimation of taste heterogeneity. 

\subsection{Constructing an RP choice dataset}
\label{sec:choice-set}
SP studies typically present respondents with a predefined set of alternatives \citep{sica2025electric,visaria2022user}. In actual travel, however, the set of PEVCSs available to a driver is session-specific and depends on travel distance, time budget, and the spatial proximity of charging stations. GPS trajectories provide an opportunity to identify a more behaviorally realistic choice set for each charging session. We therefore adopt a space-time prism \citep{hagerstrand1970people,miller1991modelling,kwan1998space,pendyala2002formulation} approach to construct session-specific PEVCS choice sets. For each charging session, $O$ and $D$ denote the observed stops immediately before and after charging, respectively, which serve as the two activity anchors defining the feasible space-time region.

The observed elapsed time between $O$ and $D$, $T_{OD}$, is used as the total time budget. For each candidate PEVCS $E$, network travel time and distance are calculated for the $O$--$E$ and $E$--$D$ legs. A station is retained in the potential path area if $t_{OE}+t_{ED} \leq T_{OD}$ and $d_{OE}+d_{ED} \leq 2d_{OD}$, where $t_{ij}$ and $d_{ij}$ denote the network travel time and distance between locations $i$ and $j$, respectively. The first constraint ensures that the station can be reached within the observed time budget, while the second requires detour distance to be no more than $d_{OD}$. A multiplier of 2 provides an intuitive geometric benchmark, with an equilateral $O$--$E$--$D$ configuration lying on the boundary of the feasible region. Increasing this multiplier would admit more spatially distant candidates, but their inclusion may remain constrained by the time budget. The difference $T_{OD}-t_{OE}-t_{ED}$ represents the time potentially available for charging and activities at each candidate station. 

For some sessions, however, the observed charging location itself falls outside the distance threshold. Such detours may arise when an activity near the charging station constitutes the main purpose of the trip, making the direct $O$--$D$ path a poor representation of the relevant choice context. For these sessions, we construct the choice set using the five PEVCSs nearest to the observed charging location.

\subsection{Joint charging-location and nearby-activity choice framework}
\label{sec:joint_choice}
We consider charging behavior as a joint decision involving both where to charge and whether to undertake a nearby activity during charging. We represent this process using a nested choice structure. At the upper branch, driver $n$ at charging session $t$ chooses a PEVCS $i$ from the session-specific choice set $\mathcal{C}_{nt}$ constructed in \cref{sec:choice-set}. Conditional on station $i$, the lower branch represents a binary choice between visiting a nearby point-of-interest (POI) and making no visit. The lower-branch expected utility is incorporated into the upper-branch charging station choice through an inclusive value reflecting place attractiveness.

\subsubsection{Upper-branch charging location choice}
\label{sec:upper}
Let $U^{loc}_{nit}=V^{loc}_{nit}+\varepsilon^{loc}_{nit}$ denote the utility that driver $n$ derives from choosing PEVCS $i$ during session $t$, where $V^{loc}_{nit}$ denotes the systematic utility and $\varepsilon^{loc}_{nit}$ is an i.i.d. Type-I extreme-value error term. The probability of driver $n$ choosing station $i$ during session $t$ is defined in \cref{eq:upper-prob}.
\begin{equation}
  P_{nt}(i) = \frac{\exp(V^{loc}_{nit})}
         {\sum_{l \in \mathcal{C}_{nt}} \exp(V^{loc}_{nlt})}
  \label{eq:upper-prob}
\end{equation}

Following \cite{potoglou2023public}, we let $V^{loc}_{nit}$ consist of trip-level attributes, station attributes, and the inclusive value from the lower branch, as shown in \cref{eq:upper-utility}. 
\begin{equation}
\begin{split}
  V^{loc}_{nit} &= \beta^{OE}_{nt} d^{OE}_{nit} + \beta^{det}_{nt}\Delta t_{nit} \\
    & + \beta^{price}_{nt} Price_{i} + \beta^{L_{2}}_{nt} Num^{L_{2}}_{i}
      + \beta^{DCFC}_{nt} Num^{DCFC}_{i} 
      + \sum_{v \in \mathcal{V}} \beta^{Net}_{v,nt} Network^v_i\\
    & + \beta^{IV}_{nt} IV_{nit}
\end{split}
\label{eq:upper-utility}
\end{equation}
where $d^{OE}_{nit}$ is the network distance from the stop preceding the charging session to candidate station $i$ (as a proxy of range anxiety), and $\Delta t_{nit}=t^{OE}_{nit}+t^{ED}_{nit}-t^{OD}_{nit}$ is the detour time required to reach station $i$. These are trip-level attributes obtained in the choice set generation procedure. 

Station-level attributes include charging price $Price_{i}$, the numbers of Level 2 and DC fast-charging ports, $Num^{L2}_{i}$ and $Num^{DCFC}_{i}$, and a set of charging-network-specific constants $Network^v_i$ characterizing the unobserved factors associated with charging networks such as ChargePoint, Tesla, and others. The last term in \cref{eq:upper-utility}, \(IV_{nit}\), summarizes the expected utility of activity opportunities available while charging at station $i$. Its parameter, $\beta^{IV}_{nt}$, reflects the scale of the lower-branch utility, capturing the sensitivity of charging location choice to the attractiveness of nearby activity opportunities. A larger positive $\beta^{IV}_{nt}$ indicates that the driver places greater weight on activities that can be undertaken while charging.

\subsubsection{Lower-level POI visitation choice}
\label{sec:lower}
Conditional on charging at station $i$, the EV driver chooses between visiting a nearby POI and making no visit. Let $r\in \{1,0\}$ denote these two alternatives, where $r=1$ represents a POI visit. The lower-level utility also consists of a systematic part and a random part, i.e., $U^{visit}_{nirt}=V^{visit}_{nirt}+\varepsilon^{visit}_{nirt}$. The systematic utility of no POI visit is normalized to zero ($V^{visit}_{ni,0,t}=0$), while the systematic utility of visiting a POI is defined in \cref{eq:lower-utility}.
\begin{equation}
\begin{split}
  V^{visit}_{ni,1,t} &= \sum_{h \in \mathcal{H}} \beta^{POI}_{h,nt} Num^{POI}_{hi} 
       + \sum_{d \in \mathcal{D}} \beta^{SV}_{d,nt} SV^d_i 
       + \beta^{\tau}_{nt} \tau_{nit}
       + ASC^{visit}_{nt}
\end{split}
\label{eq:lower-utility}
\end{equation}
where $Num^{POI}_{hi}$ denotes the number of POIs in category $h$ within a 200m buffer of station $i$. $SV^d_i$ represents the streetview perception score of station $i$, reflecting the visual characteristics of the surrounding built. Following \cite{yao2019human}'s work, we retrieve six perception dimensions from the Google Street View panorama closest to each station, including beautiful, boring, depressing, lively, safety, and wealthy. $\tau_{nit}=T^{OD}_{nt}-t^{OE}_{nit}-t^{ED}_{nit}$ is the available charging/activity time after selecting candidate station $i$, where $T^{OD}_{nt}$ is the observed time budget between the activity anchors $O$ and $D$.

In the lower-branch, the conditional probability of choosing $r$ given station $i$ and the corresponding inclusive value are defined in \cref{eq:lower-prob} and \cref{eq:IV}, respectively.
\begin{equation}
  P_{nt}(r \mid i) = \frac{\exp(V^{visit}_{nirt})}
         {\sum_{s \in \{0,1\}} \exp(V^{visit}_{nist})}
  \label{eq:lower-prob}
\end{equation}

\begin{equation}
  IV_{nit} = \ln \left[ \sum_{r \in \{0,1\}}
      \exp \left( V^{visit}_{nirt} \right) \right]
      =\ln \left[ 1+\exp \left( V^{visit}_{ni,1,t} \right) \right]
  \label{eq:IV}
\end{equation}
where the constant 1 comes from the no-visit alternative, whose systematic utility is normalized to zero. The scale of the lower-level utility is normalized to one for identification. The probability of observing a joint charging-location and POI-visitation choice is therefore defined as \cref{eq:joint-prob}.
\begin{equation}
  P_{nt}(i,r) = P_{nt}(i) P_{nt}(r \mid i)
  \label{eq:joint-prob}
\end{equation}

To account for the substantial heterogeneity in public charging behavior, we allow the taste parameters $\vect{\beta}_{nt}$ in both branches to vary across individuals $n$ and trip contexts $t$. This specification captures differences in drivers’ preferences for charging and activity attributes, while also allowing these preferences to vary across charging sessions depending on the specific travel context. The following section introduce a model for estimating $\vect{\beta}_{nt}$.

\subsection{AMXL-z specification and estimation algorithm}
\label{sec:AMXL-z}
\subsubsection{The AMXL-z model}
The joint choice framework in \cref{sec:joint_choice} allows taste parameters to vary with both the $\vect{z}$ vector and remaining unobserved heterogeneity. To estimate these two sources of variation separately, we extend the agent-based mixed logit (AMXL) model \citep{ren2022random,ren2026microtransit} to an AMXL-z specification.

For notational convenience, we compact the individual--session indices $(n,t)$ into an agent index $a=1,\ldots,A$, with each $a$ representing choice observations (or states) that share the same $\vect{z_a}$ and $\vect{\beta_a}$. For each branch $b \in \{loc,visit\}$, AMXL-z decomposes the agent-specific taste vector into three components, as shown in \cref{eq:AMXL-z-beta}.
\begin{equation}
  \vect{\beta}^{b}_{a} = \vect{\bar{\beta}}^b + g^b(\vect{z}_a) + \vect{\eta}^b_{a}, 
                       \qquad g^b(\vect{z}_a) = \bm{\Gamma}^{b\top} \bm{\phi}(\vect{z_a})
  \label{eq:AMXL-z-beta}
\end{equation}
where $\vect{\bar{\beta}}^b$ represents the population-level taste vector, $g^b(\vect{z}_a)$ captures systematic preference variation associated with observed characteristics, and $\vect{\eta}^b_{a}$ represents the remaining unobserved heterogeneity. $\bm{\phi}(\vect{z}_a)$ is a feature transformation of vector $\vect{z}$. A linear specification directly uses the original values, whereas polynomial feature expansions can accommodate nonlinear effects. $\bm{\Gamma}^{b} \in \mathbb{R}^{P \times K_b}$ is a branch-specific coefficient matrix that maps the $P$ transformed features in $\bm{\phi}(\vect{z}_a)$ to systematic shifts in the $K_b$ taste parameters. Each element $\Gamma^{b}_{pk}$ captures the contribution of feature $p$ to taste parameter $k$.

The original AMXL model can be viewed as a special case of AMXL-z in which no observed attribute enters to the vector $\vect{z}_a$. In that case, $g^b(\vect{z}_a) = 0$ and the model reduces to $\vect{\beta}^{b}_{a} = \vect{\bar{\beta}}^b + \vect{\tilde{\eta}}^b_{a}$. The limitation of the original AMXL is that the systematic variation and unobserved variation are combined in the empirical distribution of $\vect{\tilde{\eta}}^b_{a}$. AMXL-z instead explicitly separates these two sources by extending the fixed-point prior to a context-specific prior, $\vect{m}^{b}_{a} = \vect{\bar{\beta}}^b + \bm{\Gamma}^{b\top} \bm{\phi}(\vect{z_a})$, where $\vect{\bar{\beta}}^b$ serves as the population-level taste vector (or intercept), and $\bm{\Gamma}^{b\top} \bm{\phi}(\vect{z_a})$ captures taste heterogeneity associated with the $\vect{z}$ vector. The remaining deviation, $\vect{\eta}^b_{a}$, is treated as unobserved taste heterogeneity, whose distribution is learned empirically from the agent-level choice data.

Although we use a kernel-feature expansion to specify the context-specific prior in this study, the same decomposition could in principle be generalized using a more flexible function approximator, such as a neural network. Such an extension would require a different training step and is left for future research. 

Moreover, the application of AMXL-z extends beyond EV charging behavior modeling. It provides a generalizable framework for modeling taste heterogeneity using large-scale GPS data across a wide range of choice scenarios.

\subsubsection{Proposed estimation algorithm}
\label{sec:AMXL-algorithm}
Following the AMXL framework, the parameters of each branch are estimated through inverse optimization. For branch $b \in \{loc,visit\}$, let $y^b_{aiq}=1$ if agent $a$ chooses alternative $i$ on choice observation $q$, and zero otherwise. The complete inverse utility optimization problem is formulated in \crefrange{eq:amxl-obj}{eq:amxl-bounds}.
\begin{subequations}\label{eq:amxl}
\renewcommand{\theequation}{\theparentequation-\alph{equation}}
\begin{align}
  \min_{\bar{\bm{\beta}}^{b},\, \bm{\Gamma}^{b},\, \{\bm{\beta}^{b}_{a}\}_{a=1}^{A}}\quad
  & \sum_{a=1}^{A}\left[
    \begin{aligned}
      -\sum_{q \in \mathcal{T}_{a}}\;\sum_{i \in \mathcal{C}^{b}_{aq}}
        y^{b}_{aiq}\log P^{b}_{aiq}\bigl(\bm{\beta}^{b}_{a}\bigr)
        + \lambda_{b}\left\lVert \bm{\beta}^{b}_{a} - \bm{m}^{b}_{a}\right\rVert_{2}^{2}
    \end{aligned}
  \right] \label{eq:amxl-obj} \\
  \intertext{\centerline{subject to:}}
   &P^{b}_{aiq}\bigl(\bm{\beta}^{b}_{a}\bigr)
     = \frac{\exp\bigl(\bm{x}^{b\top}_{aiq}\bm{\beta}^{b}_{a}\bigr)}
           {\sum_{j \in \mathcal{C}^{b}_{aq}}
            \exp\bigl(\bm{x}^{b\top}_{ajq}\bm{\beta}^{b}_{a}\bigr)},
    \label{eq:amxl-logit} \\
   &\bm{m}^{b}_{a} = \bar{\bm{\beta}}^{b}
    + \bm{\Gamma}^{b\top}\bm{\phi}(\bm{z}_{a}), \label{eq:amxl-prior} \\
   &\bm{l}^{b} \le \bm{\beta}^{b}_{a} \le \bm{u}^{b},
    \qquad a = 1,\ldots,A. \label{eq:amxl-bounds}
\end{align}
\end{subequations}
where $\vect{x}^b_{aiq}$ contains the attributes specified in \cref{sec:joint_choice}, $\vect{\beta}^{b}_{a}$ is the corresponding agent-specific taste vector, $P^{b}_{aiq}\bigl(\bm{\beta}^{b}_{a}\bigr)$ is the probability of agent $a$ choosing alternative $i$ in observation $q$ and branch $b$, and $\bm{u}^{b}$, $\bm{l}^{b}$ are the upper and lower bounds of taste parameters, respectively. The first term in the objective function is the negative log-likelihood (NLL) of the observed choices under the agent-specific taste vector, while the second is a regularization term that penalizes deviations of the agent-specific tastes from the context-specific prior $\vect{m}_a^b$. Conditional on $\bm{\beta}^{b}_{a}$, minimizing the second term reduces to an OLS regression of the agent-specific tastes on the transformed characteristics $\bm{\phi}(\bm{z}_{a})$, with $\bar{\bm{\beta}}^{b}$ serving as the intercept.

The shrinkage parameter $\lambda_b$ determines the relative contribution of the agent-level deviations and the context-specific prior. A larger $\lambda_b$ imposes stronger shrinkage toward $\vect{m}_a^b$, forcing more of the recoverable taste variation to be represented by the systematic component $g^b(\vect{z}_a)$. A smaller $\lambda_b$ allows the agent-specific taste vector to deviate further from its prior center, leaving a larger share of heterogeneity in the residual component $\vect{\eta}_a^b$. As $\lambda_b\rightarrow\infty$, the residual deviations approach zero; as $\lambda_b\rightarrow0$, each agent's tastes are determined by its own observed choices. We therefore treat $\lambda_b$ as a hyperparameter and tune it using predictive performance on a held-out development sample. Separate values, $\lambda_{visit}$ and $\lambda_{loc}$, are selected for the lower and upper branches.

Although \cref{eq:amxl} can be solved as a single convex optimization problem, doing so is computationally demanding because it requires jointly estimating all $A$ agent-specific taste vectors together with the population parameters. We instead use a decomposition method to initialize $\vect{\bar{\beta}}^b$ and update $\bm{\Gamma}^b$ and $\vect{\beta^b_a}$ iteration by iteration. Conditional on $\vect{\bar{\beta}}^b$ and $\bm{\Gamma}^b$, the optimization decomposes into $A$ independent agent-level problems, as shown in \cref{eq:MAP}.
\begin{equation}
  \hat{\vect{\beta}}^b_a=
  \argmin_{\vect{\beta} \in \mathcal{B}_b}
    \left[
      -\sum_{q \in \mathcal{T}_{a}}\;\sum_{i \in \mathcal{C}^{b}_{aq}}
        y^{b}_{aiq}\log P^{b}_{aiq}\bigl(\bm{\beta}\bigr)
        + \lambda_{b}\left\lVert \bm{\beta} - \bm{m}^{b}_{a}\right\rVert_{2}^{2}
  \right]
  \label{eq:MAP}
\end{equation}

We formalize this decomposition in \cref{prop:decomp}. 

\begin{proposition}\label{prop:decomp}
For branch $b$ and agent a given $\lambda_b>0$, conditional on $\bar{\boldsymbol{\beta}}^b$ and $\boldsymbol{\Gamma}^b$, the problem in \crefrange{eq:amxl-obj}{eq:amxl-bounds} decomposes into $A$ independent agent-level subproblems given by \cref{eq:MAP}. Each subproblem is strictly convex and has a unique solution $\hat{\boldsymbol{\beta}}^b_a$.
\end{proposition}

\begin{proof}
Given $\bar{\boldsymbol{\beta}}^b$ and $\boldsymbol{\Gamma}^b$, the context-specific prior $\boldsymbol{m}^b_a$ in \cref{eq:amxl-prior} is a constant vector for each agent. The objective in \cref{eq:amxl-obj} can then be written as \cref{eq:prop_decomp}.
\begin{equation}\label{eq:prop_decomp}
\begin{aligned}
&\min_{\boldsymbol{\beta}^b_1,\ldots,\boldsymbol{\beta}^b_A \in \mathcal{B}_b} \; \sum_{a=1}^{A} f_a\big(\boldsymbol{\beta}^b_a\big), \\
&f_a(\boldsymbol{\beta}) = -\sum_{q\in\mathcal{T}_a}\sum_{i\in\mathcal{C}^b_{aq}} y^b_{aiq}\log P^b_{aiq}(\boldsymbol{\beta}) + \lambda_b\big\|\boldsymbol{\beta}-\boldsymbol{m}^b_a\big\|_2^2
\end{aligned}
\end{equation}
where $\mathcal{B}_b=\{\boldsymbol{\beta}\in\mathbb{R}^{K_b}:\boldsymbol{l}^b\le\boldsymbol{\beta}\le\boldsymbol{u}^b\}$ is the feasible set defined by \cref{eq:amxl-bounds}. Each $f_a$ depends only on $\boldsymbol{\beta}^b_a$, and each agent's constraint involves only its own $K_b$ parameters. The joint minimum is therefore the sum of $A$ separate minima, each given by \cref{eq:MAP}. By \cref{eq:amxl-logit}, $-\ln P^b_{aiq}(\boldsymbol{\beta}) = -\boldsymbol{x}^{b\top}_{aiq}\boldsymbol{\beta} + \ln\sum_{j\in\mathcal{C}^b_{aq}}\exp\big(\boldsymbol{x}^{b\top}_{ajq}\boldsymbol{\beta}\big)$ is a linear function plus a log-sum-exp function of $\boldsymbol{\beta}$, so the first term of $f_a$ is convex because $y^b_{aiq}\ge 0$. The second term is strictly convex for $\lambda_b>0$. Hence $f_a$ is strictly convex. Since $\mathcal{B}_b$ is non-empty, closed, bounded and convex, \cref{eq:MAP} has a unique minimizer $\hat{\boldsymbol{\beta}}^b_a$.
\end{proof}

By \cref{prop:decomp}, each subproblem involves only $K_b$ decision variables and can be solved independently across agents. With the agent-specific taste vectors fixed, the context-specific prior is updated by regressing the recovered tastes on the transformed $z$ attributes at the end of each iteration. Consistent with the nested structure, the lower branch is estimated first. The estimated lower-branch parameters are then used to compute the inclusive value for each PEVCS alternative before solving the upper-branch problem. The estimation algorithm is summarized in \cref{alg:amxl}.

\begin{algorithm}[htbp]
\caption{Decomposed estimation of the joint AMXL-z model}
\label{alg:amxl}
\small
\noindent\textbf{Input:} Lower- and upper-branch choice data; observed
characteristics $\bm{z}_{a}$; parameter bounds; candidate hyperparameter grids
$\Lambda_{visit}$ and $\Lambda_{loc}$; convergence tolerance $\epsilon$.

\begin{enumerate}[leftmargin=2.1em, labelsep=0.5em, itemsep=3pt,
                  topsep=5pt, parsep=0pt, label=\arabic*.]
  \item Divide agents into training, development, and test samples. 
  \item Estimate a 
        lower-branch MNL model using the training sample and
        initialize $\bar{\bm{\beta}}^{\mathrm{visit},(0)}$ with the MNL
        estimates and $\bm{\Gamma}^{\mathrm{visit},(0)} = \bm{0}$.
  \item For each candidate
        $\lambda_{\mathrm{visit}} \in \Lambda_{\mathrm{visit}}$:
    \begin{enumerate}[leftmargin=1.9em, labelsep=0.5em, itemsep=2pt,
                      topsep=3pt, parsep=0pt, label=\theenumi\alph*.]
      \item Compute the current context-specific prior
            $\bm{m}^{\mathrm{visit}}_{a}$ for each agent.
      \item Solve \cref{eq:MAP} independently for each agent to
            obtain $\bm{\beta}^{\mathrm{visit}}_{a}$.
      \item Update $\bar{\bm{\beta}}^{\mathrm{visit}}$ and
            $\bm{\Gamma}^{\mathrm{visit}}$ using the OLS regression.
      \item Repeat Steps 3a--3c until the relative changes in $\bm{\beta}^{\mathrm{visit}}_{a}$ and $\bm{\Gamma}^{\mathrm{visit}}$ are below $\epsilon$.
      \item Evaluate predictive negative log-likelihood on the development
            sample and select $\lambda^{*}_{\mathrm{visit}}$.
    \end{enumerate}
  \item Using the selected lower-branch model, calculate
        $\mathit{IV}_{aiq}$ for each PEVCS alternative.
  \item Add $\mathit{IV}_{aiq}$ to the upper-branch attributes; estimate an
        upper-branch MNL model and initialize
        $\bar{\bm{\beta}}^{\mathrm{loc},(0)}$ and
        $\bm{\Gamma}^{\mathrm{loc},(0)} = \bm{0}$.
  \item Repeat Steps 3a--3e over
        $\lambda_{\mathrm{loc}} \in \Lambda_{\mathrm{loc}}$ to estimate the
        upper-branch model and select $\lambda^{*}_{\mathrm{loc}}$.
  \item Recover $\hat{\bm{\eta}}^{b}_{a} = \hat{\bm{\beta}}^{b}_{a}
        - \bar{\bm{\beta}}^{b} - \bm{\Gamma}^{b\top}\bm{\phi}(\bm{z}_{a})$
        for $b \in \{\mathrm{visit}, \mathrm{loc}\}$.
\end{enumerate}

\noindent\textbf{Output:} Selected $\lambda^{*}_{\mathrm{visit}}$ and
$\lambda^{*}_{\mathrm{loc}}$, population parameters, agent-specific tastes,
systematic taste components, and residual heterogeneity for both branches.
\end{algorithm}

The proposed algorithm substantially reduces the computational burden of the full estimation problem. For each branch and a fixed $\lambda_b$, each iteration involves $A$ independent convex subproblems, each containing $K_b$ taste parameters, followed by a population-level OLS update with $P+1$ independent variables ($P$ transformed characteristics and one intercept). Because the agent-level subproblems share no decision variables, they can be solved in parallel. The estimation for different candidate values of $\lambda_b$ can also be parallelized. Compared with the self-regulated Method of Successive Averages used in the original AMXL, the OLS step directly updates the context-specific prior by solving the population-level subproblem in closed form, thereby avoiding incremental averaging and generally requiring fewer iterations to reach convergence. We further initialize the population-level tastes using the corresponding MNL estimates, which provides a closer starting point and further reduces the number of iterations. Since the estimation problem is convex for fixed $\bm{\phi}(\vect{z}_a)$ and $\lambda_b$, this initialization affects computational efficiency but not the optimal solution. Unlike MXL estimation, AMXL-z also avoids simulation draws over the taste distribution. Its working memory is primarily determined by the $A\times K_b $ matrix of agent-specific tastes and the $K_b\times(P+1)$ coefficient matrix. Finally, AMXL-z requires tuning only a small number of hyperparameters, such as $\lambda_b$ and the kernel-transformation order in $\bm{\phi}(\vect{z}_a)$, resulting in a substantially smaller tuning space than neural-network models with multiple hyperparameters.

\subsubsection{Comparison with conventional DCMs and ANN-based models}
The AMXL-z is designed to draw a balance between conventional DCMs and ANN-based models. \cref{tab:comparison} summarizes the similarities and differences among MXL, TasteNet, and AMXL-z in terms of how they represent taste heterogeneity, perform estimation, tune hyperparameters, and exploit panel choice data in prediction. 

\begin{table}[t]
\centering
\caption{Comparison between MXL, TasteNet and AMXL-z}
\label{tab:comparison}
\small
\begin{tabular}{@{}>{\raggedright\arraybackslash}p{2.6cm}>{\raggedright\arraybackslash}p{4.0cm}>{\raggedright\arraybackslash}p{4.0cm}>{\raggedright\arraybackslash}p{4.0cm}@{}}
\toprule
 & MXL & TasteNet & AMXL-z \\
\midrule
Explained heterogeneity $g(\bm{z})$ & Interaction terms specified by the analyst, usually linear & Neural network $f(\bm{z}; \bm{\omega})$ & Kernel expansion $\bm{\Gamma}^{\mathsf{T}} \bm{\phi}(\bm{z})$, estimated by OLS \\
\addlinespace
Unexplained heterogeneity $\bm{\eta}$ & Predefined parametric distribution (e.g.\ normal)  & None & Empirical distribution of the agents' estimated deviations \\
\addlinespace
Estimation & Maximum simulated likelihood with $R$ draws per individual & Gradient-based training on a non-convex function with regularization & Convex MAP problem: small agent-level problems and an OLS update \\
\addlinespace
Tuning & Number and type of draws; distribution type & Architecture, learning rate, batch size, epochs, regularization, random seed & $\lambda$ and the order of kernel transformation \\
\addlinespace
Panel choice data & Learns individual tastes from choice history through posterior updating & Contribute to training, but do not update the individual-specific residual & Learns from choice history by reweighting empirical residual taste support \\
\bottomrule
\end{tabular}
\end{table}

The first distinction concerns the representation of explained and unexplained heterogeneity. In MXL and other conventional DCMs, systematic heterogeneity is typically introduced through prespecified interactions between observed characteristics and alternative attributes, and the remaining heterogeneity is represented by a parametric distribution. TasteNet \citep{han2022neural} takes a substantially more flexible approach to explained heterogeneity by using a neural network to map observed characteristics $\vect{z}_a$ into taste parameters. However, tastes are entirely determined by $\vect{z}_a$. Although MDN-Logit \citep{li2026decomposing} introduces a mixture density network to capture unexplained heterogeneity, the validity of its assumed distributions still depends on how completely $\vect{z}_a$ captures taste variation. AMXL-z lies between these two approaches. It allows nonlinear relationships between taste heterogeneity and observed characteristics, while the remaining agent-specific deviations are retained as an empirical distribution. The optimal $\lambda^*$ reflects the extent to which heterogeneity can be explained by $\vect{z}_a$.

A second important distinction is how the models predict using panel choice data. MXL can learn from repeated choices by updating the conditional distribution of an individual's random parameters given the observed choice history \citep{train2009}. AMXL-z retains this important feature but performs the updating over an empirically recovered distribution of residual tastes. Let $s=1,\dots,S$ denote a number of agents used for simulation. The probability function can be defined as \crefrange{eq:probability-function}{eq:panel-weight}.
\begin{equation}
  \hat{P}^{b,panel}_{aiq} = \sum^S_{s=1}  w^{b}_{as} P^b_{aiq} (\vect{\beta}^{b,(s)})
  \label{eq:probability-function}
\end{equation}
\begin{equation}
  \vect{\beta}^{b,(s)} = \bar{\vect{\beta}}^b + \bm{\Gamma}^{b\top} \bm{\phi}(\vect{z_a})
                         +\hat{\vect{\eta}^b_s}
  \label{eq:panel-beta}
\end{equation}
\begin{equation}
   w^{b}_{as}
    = \frac{L^{b}_{as}}{\displaystyle\sum_{r=1}^{S} L^{b}_{ar}},
   \qquad L^{b}_{as} = \prod_{q \in \mathcal{H}^{b}_{a}} \prod_{j \in \mathcal{C}^{b}_{aq}}
    \left[ P^{b}_{ajq}\!\left( \vect{\beta}^{b,(s)} \right)
    \right]^{y^{b}_{ajq}}
  \label{eq:panel-weight}
\end{equation}
where $\vect{\beta}^{b,(s)}$ is the candidate taste vector constructed from the context-specific prior of agent $a$ and the $s$-th estimated residual, $\hat{\vect{\eta}^b_s}$. $y^b_{ajq}$ is the choice indicator. $\mathcal H^b_a$ denote the observed choice history. $L^b_{as}$ is the likelihood value measuring how well the $s$-th candidate taste vector explains the observed choice history. Without any observed choice history, all $w^{b}_{as}$ would be the same. Once panel choices become available, $S$ agents can be weighted according to how well they can explain the observed history.

In general, AMXL-z sacrifices some of the functional flexibility of a machine learning estimator in exchange for a convex and robust estimation framework, while retaining flexible residual heterogeneity and the ability to learn from panel choice histories. This trade-off is well suited to handle large-scale GPS data, which provide rich repeated behavioral observations for agent-specific estimation but often contain sociodemographic information that is too limited or noisy to justify modeling all preference heterogeneity as a flexible function of $\vect{z}$.

\section{Simulation Study}
\label{sec:simulation}
The purpose of this simulation study is to compare AMXL-z with benchmark models in terms of predictive performance, recovery of taste heterogeneity, and computational efficiency. For simplicity, we focus only on the charging location choice, using POI proximity as a proxy for lower-level activity opportunities. We generate synthetic datasets with an underlying logit model in which taste parameters vary nonlinearly with observed characteristics ($\vect{z}$) through quadratic interactions. To mimic the uncertainty and panel structure of GPS data, we introduce different levels of noise and vary the number of choice observations per individual.

\subsection{Synthetic data generation}
\label{sec:4-1}
We generate data from an MNL formulation of charging station choice with driver-specific tastes. Each driver $n$ makes $C \in \{5,10\}$ choices among $J=3$ charging stations. The systematic utility of station $i$ is given by \cref{eq:4-1-utility-function}. 
\begin{equation}
  V_{nit} = \beta_{TT,n}TT_{nit} + \beta_{POI,n}POI_{nit} - CO_{nit}
  \label{eq:4-1-utility-function}
\end{equation}
where three station attributes are considered: the travel time to the station (TT), the number of points of interest near the station (POI) and the charging cost (CO) (see \cref{tab:synthetic-z-vector} in Appendix). The $\vect{z}$ vector includes three variables: annual income (INCOME), a dummy for usually charging at night (TOD), and the average charging duration (DUR). We assign $-1$ as the ground-truth cost parameter, so that $\beta_{TT,n}$ is measured in dollars per minute of travel time and $\beta_{POI,n}$ in dollars per point of interest. Since the stations are unlabeled, the utilities contain no alternative-specific constants.

The tastes follow the decomposition of \cref{sec:lit-DCM}, $\vect{\beta}_{n} = \bar{\vect{\beta}} + g(\vect{z}_{n}) + \vect{\eta}_{n}$, with $\beta_{TT,n}$ and $\beta_{POI,n}$ specified in \crefrange{eq:beta_TT}{eq:beta_POI}. 
\begin{equation}
\begin{split}
  \beta_{TT,n} =& -0.5 - 0.2 inc_n - 0.1 tod_n + 0.15 dur_n \\
                &- 0.05 inc_n tod_n + 0.05 inc_n dur_n - 0.1tod_n dur_n + \eta_{TT,n}
  \label{eq:beta_TT}
\end{split}
\end{equation}
\begin{equation}
\begin{split}
  \beta_{POI,n} = &0.3 + 0.1 inc_n - 0.2 tod_n + 0.25 dur_n \\
                 &- 0.05 inc_n tod_n + 0.1 inc_n dur_n - 0.1tod_n dur_n + \eta_{POI,n}
  \label{eq:beta_POI}
\end{split}
\end{equation}
where driver characteristics enter the tastes in standardized form, $\vect{z_n}=(inc_n,tod_n,dur_n)$, with $inc_n=(\ln \mathrm{Income}_n -\ln 60)/0.35$, $tod_n=\mathrm{TOD}_n$, and $dur_n=(\mathrm{DUR}_n-60)/30$. The intercepts, $\vect{\bar{\beta}}=(-0.50, 0.30)^{\top}$, reflect the population-level tastes. Values of the parameters are chosen to reflect real charging preferences. For example, drivers with higher income are more sensitive to travel time and value points of interest more. Charging at night makes travel time more costly and points of interest less attractive, as many of them are closed at night (see \cref{fig:dgp_taste_function} in Appendix). We intentionally include the interactions to make the relation between the characteristics and the tastes nonlinear, thereby allowing us to test how well different models recover nonlinear relationships.

\begin{figure}[htbp] 
  \centering
  \includegraphics[width=0.85\linewidth]{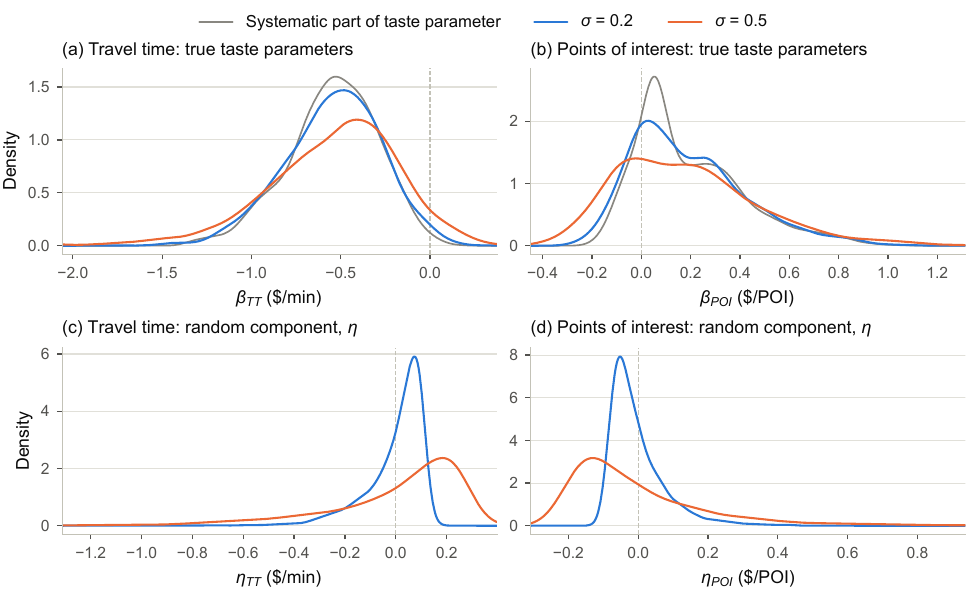}
  \caption{Distributions of the true parameters ($\vect{\beta}_n$) and the random component ($\vect{\eta}_n$)}
  \label{fig:main_taste_distribution}
\end{figure}

In addition to the systematic taste functions, we add a non-Gaussian random component to represent residual taste heterogeneity $\vect{\eta}$ that cannot be explained by $\vect{z}$. The random component is designed to be heteroskedastic and asymmetric, with its general noise level depending on $\sigma$. Details of its generation process are provided in the Appendix. The specification weakens the deterministic mapping from $\vect{z}$ to $\vect{\beta}$, which is intended to mimic large-scale GPS applications, where individual characteristics are often missing or aggregated, leaving preference variation that may be uneven across individuals and need not follow a normal distribution. We create datasets with two noise levels, $\sigma=0.2$ and $\sigma=0.5$. \cref{fig:main_taste_distribution} shows the distributions of true parameters and their random part under different noise levels. More details of individual-level parameters can be found in \cref{fig:dgp_taste_distribution} in the Appendix.

\subsection{Benchmark models and experimental setup}
\label{sec:4-2}

\subsubsection{Models in comparison}
\label{sec:4-2-1}
We compare AMXL-z with three benchmarks: an MNL and an MXL, which represent conventional DCMs, and TasteNet-MNL \citep{han2022neural}, which represents ANN-based models. All models use the same utility function of the data-generating process, so they differ only in how the heterogeneity in $\beta_{TT,n}$ and $\beta_{POI,n}$ is estimated. INCOME, TOD and DUR are treated as $z$ variables.

For the MNL benchmark, $\beta_{TT,n}$ and $\beta_{POI,n}$ are assumed to be fixed across individuals. For the MXL benchmark, both taste parameters are specified as normally distributed random parameters. To allow the two DCMs to capture the impacts of driver characteristics, we include interactions between each of the three $z$ variables and the two taste parameters. The coefficients associated with these interaction terms are treated as fixed parameters in both MNL and MXL. These models are estimated using the Python xlogit package.

For TasteNet-MNL, the three $z$ variables are fed into a neural network that directly maps $z$ variables to the two taste parameters, $\beta_{TT,n}$ and $\beta_{POI,n}$. The resulting individual-specific tastes are then embedded in a MNL utility function for prediction. Following \cite{han2022neural}, the neural network has one hidden layer of $h$ ReLU unites and a linear output layer. The model is trained using a PyTorch package in Python by minimizing the negative log-likelihood of the observed choices with a stochastic gradient-based optimization algorithm (Adam) and an L2 penalty on the weights (weight decay). 

For AMXL-z, we treat each driver as an agent and consider two specifications. AMXL-linear uses a linear transformation of the z variables, whereas AMXL-quadratic uses a quadratic transformation that additionally captures nonlinear effects and interactions. We therefore introduce two vectors: $\bm{\phi}_L(\vect{z})=(inc, tod, dur)^{\top}$ and $\bm{\phi}_Q(\vect{z})=(inc, tod, dur, inc^2, tod^2, dur^2, inc \times tod, inc \times dur, tod \times dur)^{\top}$. The models are estimated in Python using the decomposition algorithm described in \cref{sec:AMXL-algorithm}. The population-level taste vector is initialized using the corresponding MNL estimates, while the interaction terms are initialized at zero. Iteration stops when the relative change in $\vect{m_a}$ (see \cref{eq:amxl-obj}) falls below 0.01\%. 

\subsubsection{Experimental design}
\label{sec:4-2-2}
The experiments use four synthetic datasets defined by the combination of two noise levels, $\sigma=0.2$ and $\sigma=0.5$, and two panel sizes, $C=5$ and $C=10$. For each dataset, the 2,000 drivers are randomly split into training, development, and test sets in a 60:20:20 ratio. The 1,200 training drivers are used to estimate model parameters, the 400 development drivers are used for hyperparameter selection, and the remaining 400 test drivers are for out-of-sample evaluation. A summary of taste parameters across the training, development, and test samples can be found in \cref{tab:train_dev_test_split} in the Appendix.

For prediction, each driver is assigned five additional choices that are excluded from model estimation. In other words, the first $C$ choices are used for estimation and the subsequent five are reserved exclusively for evaluating predictive performance. Predictive performances are evaluated in the three settings: in-sample, for the training drivers; out-of-sample without panel data, for the test drivers predicted from $\vect{z}_n$ alone; and out-of-sample with panel, for the test drivers predicted from $\vect{z}_n$ and $\vect{\eta}_n$ weighted by $C$ history choices.

For TasteNet, we search over the number of hidden units, $h \in \{50,100,200\}$, and the weight decay, $\{10^{-4},10^{-3},10^{-2}\}$, and train each of the nine combinations ten times from different random initializations, which gives 90 trials per dataset. For AMXL-linear and AMXL-quadratic, the parameter $\lambda$ is selected from the grid $\{10^{-5},10^{-4},...,10^{5}\}$ with ten values. The best hyperparameter configuration for each model is selected based on the negative log-likelihood (NLL) on the development set, with the NLL calculation described in the following subsection.

\subsubsection{Performance metrics}
\label{sec:4-2-3}
Let $\mathcal{S}$ denote the choices for evaluation, let $J=3$ denote the number of alternatives, and let $\hat{P}_{nit}$ be the predicted probability that driver $n$ chooses station $i$ on observation $t$. We measure predictive performance by the NLL per choice, prediction accuracy (ACC), and McFadden's $\rho^2$, as shown in \crefrange{eq:NLL}{eq:rho-square}.
\begin{equation}
\mathrm{NLL} = -\frac{1}{J|\mathcal{S}|} \sum_{(n,t)\in\mathcal{S}} \sum_{i=1}^{J}
                y_{nit}\ln \hat{P}_{nit}
    \label{eq:NLL}
\end{equation}
\begin{equation}
    \mathrm{ACC} = \frac{1}{|\mathcal{S}|} \sum_{(n,t)\in\mathcal{S}} y_{ni_{nt}^{*}t},
    \qquad i_{nt}^{*} = \arg\max_{i}\hat{P}_{nit}
    \label{eq:ACC}
\end{equation}
\begin{equation}
    \rho^{2} = 1- \frac{\sum_{(n,t)\in\mathcal{S}} \sum_{i} y_{nit}\ln \hat{P}_{nit}}
    {|\mathcal{S}|\ln(1/J)}
    \label{eq:rho-square}
\end{equation}

Taste recovery is measured by the root mean squared error (RMSE) between the predicted and the true taste parameters, as shown in \cref{eq:RMSE}.
\begin{equation}
    \mathrm{RMSE}_{k} = \sqrt{\frac{1}{N}\sum_{n=1}^{N} 
                        \left(\hat{\beta}_{k,n}-\beta_{k,n}\right)^2},
                        \qquad k\in\{TT,POI\}
    \label{eq:RMSE}
\end{equation}

Computational efficiency is evaluated on a local machine equipped with an Intel Core i7-10875H CPU and 32GB of RAM. We record the total CPU time required to estimate each model, including hyperparameter tuning, together with peak RAM usage during estimation. To facilitate comparison across models with different numbers of hyperparameter configurations, we also report the average CPU time per model fit.

\subsection{Experimental results}
\label{sec:4-3}
\subsubsection{Predictive performance}
\label{sec:4-3-1}
Model predictive performance is evalutated using NLL, ACC, and McFadden's $\rho^2$ introduced in \cref{sec:4-2-3}. Lower NLL and higher ACC/$\rho^2$ indicate better predictive performance. \cref{tab:section4_prediction_accuracy} summarizes the in-sample and out-of-sample results across different noise levels $\sigma$ and observations per driver $C$, with out-of-sample predictions further distinguished between no-panel and panel settings.

\begin{table}[htbp]
\centering
\footnotesize
\caption{Prediction accuracy. Bold: best model in each column of a block; italics: the true tastes (the ceiling).}
\label{tab:section4_prediction_accuracy}
\resizebox{\linewidth}{!}{%
\begin{tabular}{lccccccccc}
\toprule
 & \multicolumn{3}{c}{In-sample} & \multicolumn{3}{c}{Out-of-sample (no panel)} & \multicolumn{3}{c}{Out-of-sample (panel)} \\
\cmidrule(lr){2-4}\cmidrule(lr){5-7}\cmidrule(lr){8-10}
Model & NLL & ACC & $\rho^2$ & NLL & ACC & $\rho^2$ & NLL & ACC & $\rho^2$ \\
\midrule
\multicolumn{10}{l}{\textit{$\sigma = 0.2$, $C = 5$}} \\
MNL & 0.174 & 0.782 & 0.524 & 0.180 & 0.774 & 0.510 & 0.180 & 0.774 & 0.510 \\
MXL & 0.169 & 0.786 & 0.539 & 0.178 & 0.774 & 0.514 & 0.173 & 0.782 & 0.528 \\
TasteNet & 0.164 & \textbf{0.795} & 0.552 & \textbf{0.169} & 0.785 & \textbf{0.539} & 0.169 & 0.785 & 0.539 \\
AMXL-linear & 0.170 & 0.783 & 0.537 & 0.179 & 0.775 & 0.513 & 0.175 & 0.781 & 0.523 \\
AMXL-quadratic & \textbf{0.163} & 0.794 & \textbf{0.556} & \textbf{0.169} & \textbf{0.787} & \textbf{0.539} & \textbf{0.167} & \textbf{0.790} & \textbf{0.543} \\
\textit{True tastes} & \textit{0.153} & \textit{0.810} & \textit{0.583} & \textit{0.155} & \textit{0.809} & \textit{0.578} & \textit{0.155} & \textit{0.809} & \textit{0.578} \\
\addlinespace
\multicolumn{10}{l}{\textit{$\sigma = 0.2$, $C = 10$}} \\
MNL & 0.174 & 0.782 & 0.524 & 0.180 & 0.774 & 0.508 & 0.180 & 0.774 & 0.508 \\
MXL & 0.166 & 0.787 & 0.546 & 0.179 & 0.776 & 0.512 & 0.171 & 0.785 & 0.533 \\
TasteNet & 0.164 & \textbf{0.796} & 0.552 & \textbf{0.169} & 0.786 & \textbf{0.538} & 0.169 & 0.786 & 0.538 \\
AMXL-linear & 0.166 & 0.788 & 0.546 & 0.179 & 0.776 & 0.512 & 0.171 & 0.790 & 0.532 \\
AMXL-quadratic & \textbf{0.161} & 0.795 & \textbf{0.560} & \textbf{0.169} & \textbf{0.787} & \textbf{0.538} & \textbf{0.166} & \textbf{0.792} & \textbf{0.547} \\
\textit{True tastes} & \textit{0.153} & \textit{0.810} & \textit{0.583} & \textit{0.155} & \textit{0.809} & \textit{0.578} & \textit{0.155} & \textit{0.809} & \textit{0.578} \\
\addlinespace
\multicolumn{10}{l}{\textit{$\sigma = 0.5$, $C = 5$}} \\
MNL & 0.202 & 0.743 & 0.447 & 0.220 & 0.728 & 0.399 & 0.220 & 0.728 & 0.399 \\
MXL & 0.176 & 0.771 & 0.520 & 0.209 & 0.728 & 0.430 & 0.186 & 0.751 & 0.493 \\
TasteNet & 0.193 & 0.756 & 0.473 & 0.210 & \textbf{0.741} & 0.427 & 0.210 & 0.741 & 0.427 \\
AMXL-linear & 0.177 & 0.770 & 0.517 & 0.208 & 0.728 & 0.432 & 0.185 & 0.750 & 0.494 \\
AMXL-quadratic & \textbf{0.175} & \textbf{0.773} & \textbf{0.523} & \textbf{0.200} & 0.733 & \textbf{0.455} & \textbf{0.182} & \textbf{0.753} & \textbf{0.502} \\
\textit{True tastes} & \textit{0.148} & \textit{0.812} & \textit{0.596} & \textit{0.147} & \textit{0.818} & \textit{0.597} & \textit{0.147} & \textit{0.818} & \textit{0.597} \\
\addlinespace
\multicolumn{10}{l}{\textit{$\sigma = 0.5$, $C = 10$}} \\
MNL & 0.202 & 0.743 & 0.448 & 0.220 & 0.727 & 0.398 & 0.220 & 0.727 & 0.398 \\
MXL & 0.168 & 0.784 & 0.542 & 0.208 & 0.729 & 0.431 & 0.175 & 0.768 & 0.523 \\
TasteNet & 0.193 & 0.758 & 0.474 & 0.209 & \textbf{0.740} & 0.430 & 0.209 & 0.740 & 0.430 \\
AMXL-linear & 0.168 & 0.784 & 0.541 & 0.208 & 0.727 & 0.433 & 0.173 & 0.772 & 0.529 \\
AMXL-quadratic & \textbf{0.166} & \textbf{0.787} & \textbf{0.546} & \textbf{0.199} & 0.736 & \textbf{0.456} & \textbf{0.170} & \textbf{0.779} & \textbf{0.535} \\
\textit{True tastes} & \textit{0.148} & \textit{0.812} & \textit{0.596} & \textit{0.147} & \textit{0.818} & \textit{0.597} & \textit{0.147} & \textit{0.818} & \textit{0.597} \\
\bottomrule
\end{tabular}}
\end{table}

Overall, AMXL-quadratic consistently achieves the lowest NLL among all estimated models across the four experimental settings, demonstrating its superior predictive performance. Under relatively low noise $\sigma=0.2$, TasteNet achieves marginally higher accuracy both in-sample and out-of-sample-accuracy without panel information. For out-of-sample prediction with panel information, AMXL-quadratic generally outperforms the other models. For example, when $\sigma=0.2$ and $C=10$, AMXL-quadratic achieves an NLL of 0.166 and an ACC of 0.792, compared with 0.171 and 0.785 for MXL, and 0.169 and 0.786 for TasteNet. The performance differences become more pronounced as the noise level increases to $\sigma=0.5$. While TasteNet maintains a slight advantage in out-of-sample ACC under the no-panel setting, AMXL-quadratic consistently achieves lower NLL and higher $\rho^2$. Under the panel setting with $C=10$, AMXL-quadratic achieves an NLL of 0.170, compared with 0.175 for MXL and 0.209 for TasteNet.

Although AMXL-linear generally underperforms AMXL-quadratic, the performance gap narrows as the noise level increases. For example, when $C=10$, the difference in panel NLL between the two specifications decreases from 0.005 at $\sigma=0.2$ to 0.002 at $\sigma=0.5$. Under higher noise level, AMXL-linear also begins to outperform TasteNet. In the panel setting with $C=10$, AMXL-linear achieves an NLL of 0.173 and an ACC of 0.772, compared with 0.209 and 0.740 for TasteNet. These findings suggest that while a more flexible nonlinear specification generally contributes to predictive gains, the simpler linear specification remains competitive. This also highlights the potential value of adopting a more parsimonious functional form when the expected noise level is higher.

\subsubsection{Recovery of taste heterogeneity}
\label{sec:4-3-2}
\begin{figure}[htbp]
  \centering
  \includegraphics[width=\linewidth]{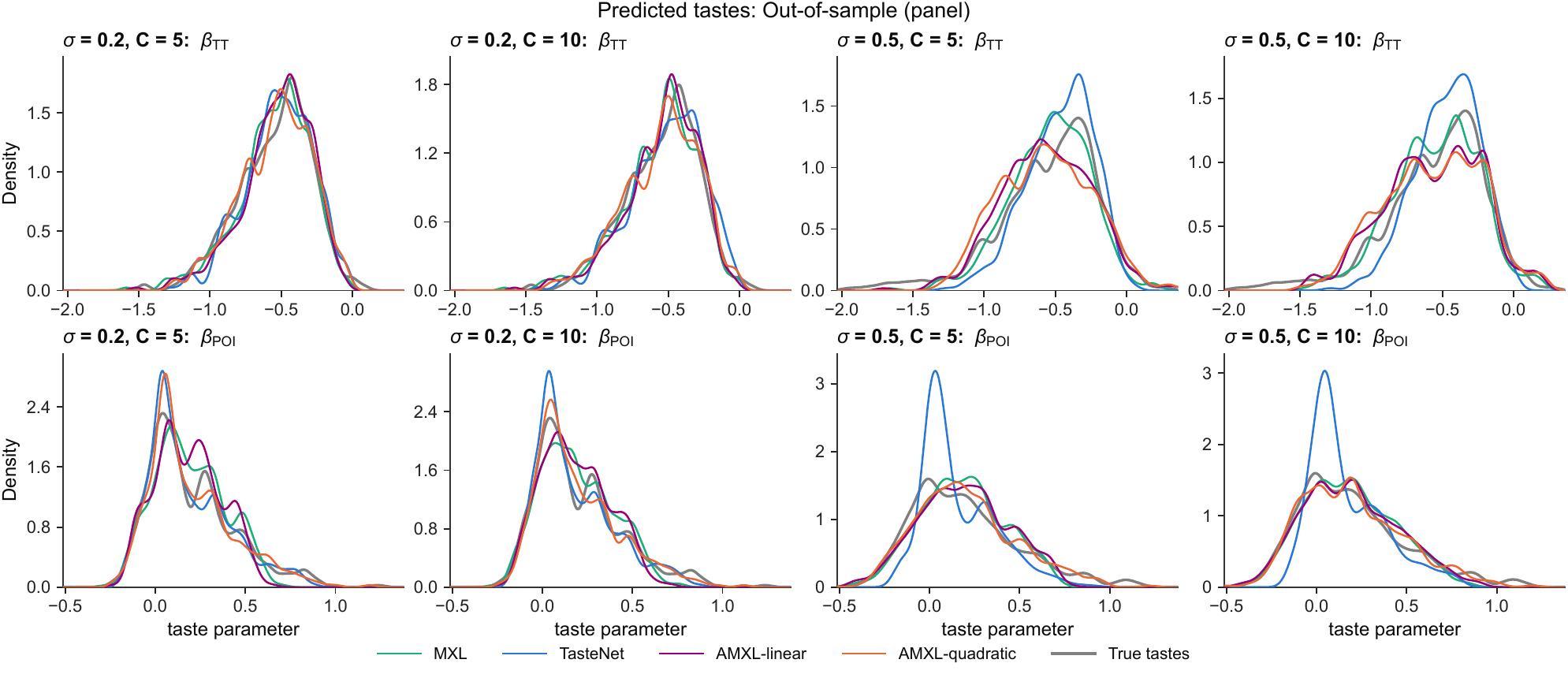}
  \caption{True tastes of all 2000 drivers against annual income. Points: individual tastes; black markers: means within income deciles; black line: population mean taste.}
  \label{fig:distributions_short}
\end{figure}

Parameter recovery is evaluated using the RMSE between the estimated and true individual taste parameters. Lower RMSE indicates more accurate recovery of taste heterogeneity. \cref{tab:section4_parameter_recovery} reports the results across different values of $\sigma$ and $C$, while \cref{fig:distributions_short} compares the distributions of recovered tastes with their true distributions.

\begin{table}[htbp]
\centering
\footnotesize
\caption{Parameter recovery: RMSE between the predicted and the true individual tastes. Bold: best model in each column of a block.}
\label{tab:section4_parameter_recovery}
\begin{adjustbox}{max width=\linewidth}
\begin{tabular}{lcccccc}
\toprule
 & \multicolumn{2}{c}{In-sample} & \multicolumn{2}{c}{Out-of-sample (no panel)} & \multicolumn{2}{c}{Out-of-sample (panel)} \\
\cmidrule(lr){2-3}\cmidrule(lr){4-5}\cmidrule(lr){6-7}
Model & $\beta_{TT}$ & $\beta_{POI}$ & $\beta_{TT}$ & $\beta_{POI}$ & $\beta_{TT}$ & $\beta_{POI}$ \\
\midrule
\multicolumn{7}{l}{\textit{$\sigma = 0.2$, $C = 5$}} \\
MNL & 0.143 & 0.123 & 0.143 & 0.129 & 0.143 & 0.129 \\
MXL & 0.131 & 0.104 & 0.142 & 0.124 & 0.133 & 0.106 \\
TasteNet & 0.121 & 0.080 & 0.125 & \textbf{0.085} & 0.125 & 0.085 \\
AMXL-linear & 0.131 & 0.111 & 0.141 & 0.126 & 0.132 & 0.115 \\
AMXL-quadratic & \textbf{0.111} & \textbf{0.076} & \textbf{0.122} & 0.087 & \textbf{0.116} & \textbf{0.080} \\
\addlinespace
\multicolumn{7}{l}{\textit{$\sigma = 0.2$, $C = 10$}} \\
MNL & 0.143 & 0.125 & 0.144 & 0.132 & 0.144 & 0.132 \\
MXL & 0.128 & 0.093 & 0.147 & 0.125 & 0.129 & 0.097 \\
TasteNet & 0.118 & 0.082 & 0.124 & 0.088 & 0.124 & 0.088 \\
AMXL-linear & 0.123 & 0.098 & 0.144 & 0.127 & 0.123 & 0.103 \\
AMXL-quadratic & \textbf{0.108} & \textbf{0.067} & \textbf{0.122} & \textbf{0.086} & \textbf{0.111} & \textbf{0.073} \\
\addlinespace
\multicolumn{7}{l}{\textit{$\sigma = 0.5$, $C = 5$}} \\
MNL & 0.307 & 0.222 & 0.308 & 0.235 & 0.308 & 0.235 \\
MXL & 0.253 & 0.156 & 0.301 & 0.225 & \textbf{0.255} & 0.166 \\
TasteNet & 0.297 & 0.199 & 0.303 & 0.212 & 0.303 & 0.212 \\
AMXL-linear & 0.257 & 0.156 & 0.302 & 0.225 & 0.256 & 0.165 \\
AMXL-quadratic & \textbf{0.252} & \textbf{0.148} & \textbf{0.293} & \textbf{0.209} & \textbf{0.255} & \textbf{0.160} \\
\addlinespace
\multicolumn{7}{l}{\textit{$\sigma = 0.5$, $C = 10$}} \\
MNL & 0.308 & 0.223 & 0.309 & 0.236 & 0.309 & 0.236 \\
MXL & 0.225 & 0.129 & 0.301 & 0.225 & \textbf{0.226} & 0.139 \\
TasteNet & 0.297 & 0.197 & 0.301 & 0.211 & 0.301 & 0.211 \\
AMXL-linear & 0.229 & 0.124 & 0.304 & 0.225 & 0.227 & 0.131 \\
AMXL-quadratic & \textbf{0.224} & \textbf{0.118} & \textbf{0.293} & \textbf{0.209} & 0.228 & \textbf{0.126} \\
\bottomrule
\end{tabular}
\end{adjustbox}
\end{table}

Overall, AMXL-quadratic achieves the lowest RMSE in most settings, while the relative performance of MXL, TasteNet, and AMXL-linear varies with noise levels and panel information. Under low noise $\sigma=0.2$, TasteNet performs relatively well. However, its performance deteriorates substantially as noise increases. For example, when $C=5$, its panel RMSE increases from 0.125 to 0.303 for $\beta_{TT}$ and from 0.085 to 0.212 for $\beta_{POI}$. \cref{fig:distributions_short} further shows that TasteNet produces an overly concentrated distribution around zero for $\beta_{POI}$ under high noise, failing to recover the true taste distribution. This suggests that TasteNet is more susceptible to overfitting noisy observations.

Increasing the number of observations per individual improves parameter recovery for both MXL and AMXL, particularly when panel information is available. At $\sigma=0.5$, increasing $C$ from 5 to 10 reduces the panel RMSE of $\beta_{POI}$ from 0.166 to 0.139 for MXL, from 0.165 to 0.131 for AMXL-linear, and from 0.160 to 0.126 for AMXL-quadratic. In contrast, TasteNet's RMSE remains nearly unchanged (0.212 versus 0.211), as it lacks a mechanism to update individual taste estimates using repeated choices. Together, these effects explain why even AMXL-linear consistently outperforms TasteNet under high noise, despite its simpler functional specification.

Notably, MXL also becomes more competitive when both $\sigma$ and $C$ increase. At $\sigma=0.5$ and $C=10$, MXL achieves a slightly lower panel RMSE for $\beta_{TT}$ than AMXL-quadratic (0.226 versus 0.228). This may reflect two factors: greater noise makes the underlying taste distribution more closely approximated by MXL's assumption on normal distribution, while additional panel observations improve the estimation of individual random parameters. However, the computational cost of MXL also increases with $C$, as discussed in the following section.

\subsubsection{Computational efficiency}
\label{sec:4-3-3}
Computational efficiency is evaluated using CPU time and peak RAM usage during model estimation. \cref{tab:section4_computational_time} summarizes the results for 1,200 training drivers under different values of $\sigma$ and $C$. The number of model fits is also reported to account for differences in estimation procedures.

Although MXL requires shorter CPU time, its memory consumption increases substantially with the number of observations per individual. When $C$ increases from 5 to 10, MXL's peak RAM usage rises from 1,099 MB to 1,833 MB. This raises concerns about the scalability of MXL estimation to large-scale mobility datasets, where both the number of individuals and repeated observations can be substantially larger.

In contrast, AMXL adopts an agent-level estimation framework that decomposes the estimation problem into individual-level contributions, allowing observations to be processed in "batches", similar to mini-batch gradient descent in neural networks. This avoids the need to process the entire dataset simultaneously and reduces memory requirements. As shown in \cref{tab:section4_computational_time}, peak RAM usage remains nearly constant as $C$ doubles. Both AMXL-linear and AMXL-quadratic also require substantially less CPU time than TasteNet. This is because the relatively small number of hyperparameters in AMXL reduces the computational time required for hyperparameter tuning. These results highlight the memory efficiency and scalability potential of AMXL, making it particularly suitable for estimating heterogeneous choice behavior from large-scale mobility data.

\begin{table}[htbp]
\centering
\footnotesize
\setlength{\tabcolsep}{4pt}
\caption{Computation time and peak RAM usage of model training/estimation (1,200 training drivers).}
\label{tab:section4_computational_time}
\begin{adjustbox}{max width=\linewidth}
\begin{tabular}{lrrrrrrr}
\toprule
 & & \multicolumn{2}{c}{CPU time (s), $\sigma = 0.2$} & \multicolumn{2}{c}{CPU time (s), $\sigma = 0.5$} & \multicolumn{2}{c}{Peak memory (MB)} \\
\cmidrule(lr){3-4}\cmidrule(lr){5-6}\cmidrule(lr){7-8}
Model & Fits & $C = 5$ & $C = 10$ & $C = 5$ & $C = 10$ & $C = 5$ & $C = 10$ \\
\midrule
MNL & 1 & 3.0 & 2.9 & 2.8 & 2.8 & 226 & 233 \\
MXL & 1 & 5.6 & 8.7 & 6.8 & 12.3 & 1,099 & 1,833 \\
TasteNet & 90 & 525 & 869 & 425 & 630 & 465 & 462 \\
AMXL-linear & 11 & 133 & 143 & 123 & 143 & 367 & 367 \\
AMXL-quadratic & 11 & 200 & 173 & 144 & 166 & 367 & 372 \\
\bottomrule
\end{tabular}
\end{adjustbox}
\end{table}

\section{Case Study in the San Francisco Bay Area}
\label{sec:case}

\subsection{Data collection}
\label{sec:data_collection}

\subsubsection{Public charging station data}
\label{sec:data_charing_station}
\begin{figure}[htbp]
  \centering
  \includegraphics[width=\linewidth]{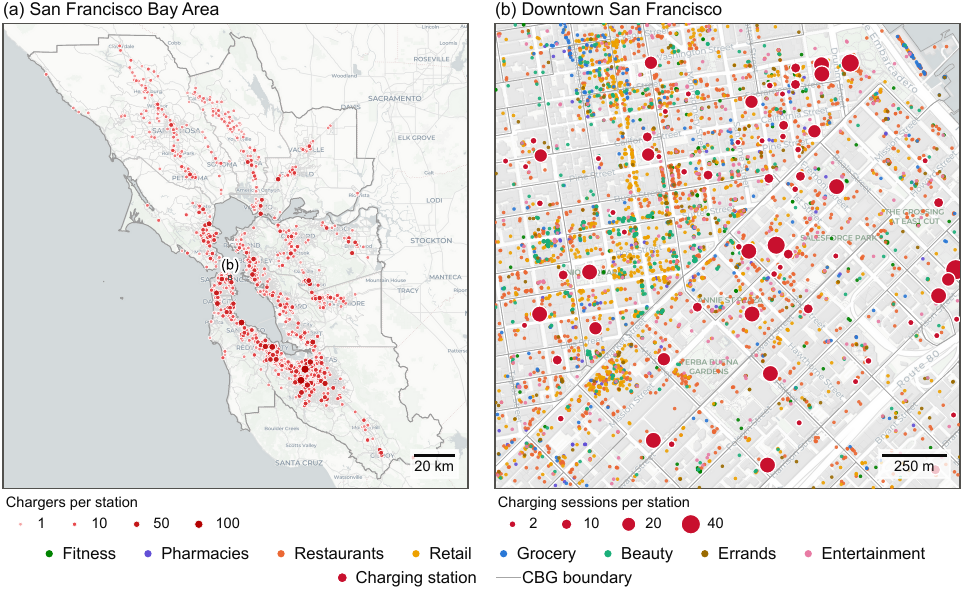}
  \caption{Study area and data example. (a) The San Francisco Bay Area and the 1,961 public charging stations in the EVCS database. Marker size increase with the number of chargers at a station. (b) Downtown San Francisco area. Marker size increases with the number of observed charging sessions at the station.}
  \label{fig:bay-area}
\end{figure}

Public EV charging station (PEVCS) data are compiled from three sources: the U.S. Department of Energy's Alternative Fuels Data Center (AFDC) \citep{afdc2025}, Open Charge Map \citep{ocm2025}, and SafeGraph POI data \citep{safegraph2022}. Records from the three sources are spatially matched and deduplicated to construct a comprehensive port-level charging dataset for the San Francisco Bay Area. The resulting dataset contains 6,761 charging ports. We further aggregate nearby ports into station-level locations using the DBSCAN algorithm \citep{khan2014dbscan}. Using a 50-m neighborhood threshold, ports within the same spatial cluster are treated as a single station. This procedure consolidates the 6,761 charging ports into 1,961 public charging stations, as shown in \cref{fig:bay-area}-(a).

Available PEVCS attributes include number of chargers per charger type (Level 2 or DCFC), charging network affiliation (ChargePoint, Tesla, other networks, and unknown), and charging price (\$/kWh, excluding parking fees). Because charging price attribute contains substantial missing values, we impute missing prices using observed prices from comparable stations. Specifically, PEVCSs in the Bay Area are grouped by charger type and charging network affiliation, and the mean observed price within each group is assigned to stations with missing price information.

\subsubsection{POI and panorama data}
\label{sec:data_POI_panorama}
We create our POI dataset from a subset of SafeGraph data, including eight POI categories: fitness, pharmacies, restaurants, retail, grocery, beauty, errands, and entertainment. The number of POIs in each category within a 200m buffer of each charging station is included in the lower-level utility function (\cref{eq:lower-utility}).

The $360^\circ$ street-view panoramas are collected using the Street View Download 360 tool (svd360.com), with each panorama retrieved from Google Street View (GSV) based on the geolocation of a PEVCS. We then use the open-source tool developed by \cite{yao2019human} to extract six visual perception dimensions from each panorama: beautiful, boring, depressing, lively, safety, and wealthy. Each dimension is measured on a 0--100 scale and is included as a separate attribute in \cref{eq:lower-utility} to characterize how drivers perceive the built environment surrounding a charging station.

Across the 1,961 stations, the mean scores (standard deviations) are 62.69 (5.57) for beautiful, 33.61 (8.03) for boring, 56.94 (5.63) for depressing, 33.31 (7.47) for lively, 37.70 (4.89) for safety, and 51.06 (4.67) for wealthy. Among the six dimensions, beautiful has the highest average score, while lively and boring have the lowest, with boring and lively exhibiting relatively greater variation across PEVCS surrounding environments. Examples of station panoramas and corresponding visual perception scores are provided in \cref{fig:gsv_examples} in the Appendix.

\subsubsection{GPS data}
\label{sec:data_GPS}
Our model builds on the public EV charging sessions and nearyby POI visits identified by \cite{clark2026using} from anonymized Cuebiq mobile phone data \citep{cuebiq2026} in San Francisco Bay Area over March–May 2022. 
Cuebiq is a location data intelligence company that collects anonymous, privacy-compliant location data of mobile devices using their software development kit (SDK) technology in mobile applications and privacy framework. Cuebiq processes data collected from mobile devices whose owners have actively opted in to share their location, and requires all application partners to disclose their relationship with Cuebiq, directly or by category, in the privacy policy. With this commitment to privacy, the dataset contains location data for roughly 15 million daily active users in the United States. Individual-level data analysis was done only within Cuebiq's Data Platform. All data analyzed in this study are aggregated to preserve privacy. Cuebiq mobility data are available through the Social Impact program: \url{https://cuebiq.com/demo}. 
Mobile phone users with a median of at least 36 location pings per day were retained, resulting in a sample of 232,662 individuals. 

Charging sessions were identified from low-speed GPS observations near public charging stations \citep{clark2026using}. Specifically, a candidate event began with a location ping moving below 10 km/h within 10 m of a charging station. Neighboring observations were then traced backward and forward until a speed or distance threshold was exceeded. A charging session was retained when at least two observations fell within 10 m of the station and the inferred stay lasted at least 10 minutes. A subset of inferred EV drivers were identified using behavioral patterns such as gas-station visits, vehicle miles traveled, the number of unique charging stations visited, and repeated charging stops. The processed dataset includes 8,930 public charging sessions made by 1,747 inferred EV drivers and was validated using EV Watts charging records \citep{EVWatts2025}. For each charging session, the observed stops immediately before and after the charging stop, together with the elapsed time between them, define the space-time constraints used to construct the PEVCS choice set. On average, each observed charging session has about 85 PEVCS alternatives, which is substantially more than the 2--5 alternatives assumed in SP surveys. 

\cite{clark2026using} also identified activities undertaken during each charging session. Cuebiq-detected stops within 250 m of the charging station were examined, and a POI visit was assigned when the stop was within 50 m of a SafeGraph POI centroid or intersected its polygon. POIs were grouped into eight categories, including fitness, pharmacies, restaurants, retail, grocery, beauty, errands, and entertainment. Medical and educational POIs were excluded for privacy reasons. A total of 1,634 POI visits were identified during charging sessions. \cref{fig:bay-area}-(b) visualizes PEVCSs and POIs in downtown San Francisco, in which the station marker size reflects the number of observed charging sessions. Individual GPS trajectories are not shown for privacy reasons.

The GPS data allow us to consider two groups of z variables. The first consists of sociodemographic attributes associated with each phone user’s inferred home location at the census-block-group (CBG) level. Based on their potential relevance to charging behavior and spatial variation across CBGs, we include the share of EV drivers (the number of identified EV drivers divided by the total number of observed phone users), and median household income (obtained from census data). The second group captures charging session context. Using the timestamp of each observed charging session, we construct binary indicators for whether the session occurs on a weekend, during daytime (8:00 a.m.–8:00 p.m.), or during the morning peak (8:00–10:00 a.m.). We also include the trip length of a charging session (network distance (km) between stops immediately before and after the charging session).

\subsection{Model results}
\label{sec:model_results}
We treat each charging session as an agent, allowing taste parameters to vary across both EV drivers and charging contexts. We report results from the AMXL-linear specification because it outperforms in the quadratic specification in out-of-sample accuracy. Due to data privacy restrictions, all models are estimated within the Spectus data platform (platform.spectus.ai). We use MNL as the benchmark for practical computational reasons: MXL estimation exceeds the platform's 12 GB RAM limit, while TasteNet cannot be reliably implemented because of package compatibility issues and the platform's execution-time constraints for long-running tasks. The full AMXL-z estimation converged in 29.76 minutes on the platform. \cref{tab:variable_definitions} in the Appendix summarizes the upper-level and lower-level alternative attributes used in the model, and the z variables entering the AMXL-z specification.

\subsubsection{Prediction accuracy}
\label{sec:result_prediction_accuracy}
\begin{figure}[htbp]
  \centering
  \includegraphics[width=\linewidth]{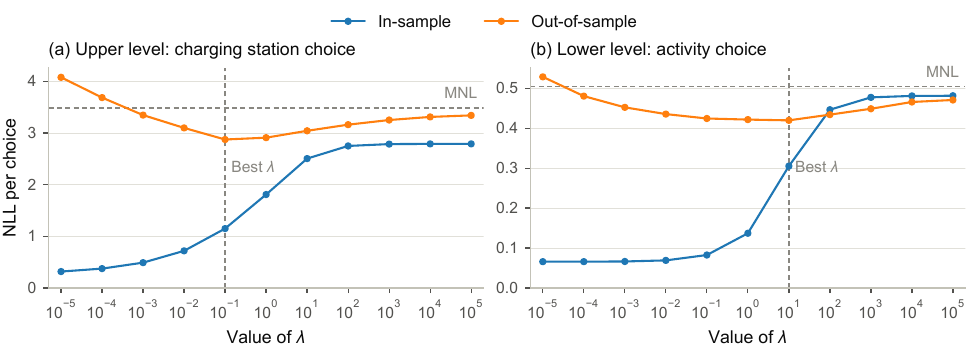}
  \caption{In-sample and out-of-sample NLL across values of $\lambda$.}
  \label{fig:prediction_accuracy}
\end{figure}

\cref{fig:prediction_accuracy} shows how predictive performance changes with the agent-level deviation penalty, $\lambda$. At the selected $\lambda$, AMXL-z achieves lower out-of-sample NLL than the MNL benchmark at both levels, indicating that allowing agent-specific taste deviations provides additional predictive value beyond population-level effects and their interactions with z variables. As $\lambda$ becomes sufficiently large, both the in-sample and out-of-sample NLL converge toward stable values. This is because under a very strong penalty, agent-level deviations are increasingly suppressed, and the model approaches an MNL specification with taste heterogeneity captured primarily through interactions with z variables.

The upper-level charging station choice exhibits larger NLL per choice than the lower-level POI visitation choice. The difference reflects the larger choice set at the upper level, where drivers choose among an average of 85 candidate stations, compared with the binary visit/no-visit decision at the lower level. The upper level also selects a smaller $\lambda$ ($10^{-1}$) than the lower level ($10^{1}$), implying a larger proportion of empirically learned agent-level deviations. This pattern suggests that, as the choice context becomes more complex, a larger share of taste heterogeneity remains unexplained by the z variables and therefore need to be learned directly from observed choice behavior.

\subsubsection{Population tastes and z effects}
\label{sec:result_basic_statistics}
\cref{tab:upper_level_results} and \cref{tab:lower_level_results} report the population-level taste estimates and their interactions with the context variables for the upper- and lower-level choices, respectively. The intercept in each row corresponds to the population taste $\bar{\beta}$. Each subsequent z coefficient represents the change in that taste associated with a one-standard-deviation increase in the corresponding context variable. The adjusted $R^2$ reports the goodness of fit from regressing the estimated agent-level tastes on z variables ($\bm{\Gamma}^b$ in the final iteration). A high adjusted $R^2$ suggests that the corresponding taste varies primarily with z variables, whereas a low value indicates that a larger share of its heterogeneity remains in the empirically learned agent-level deviation. The parameters and standard errors (in parentheses) for some attributes have been multiplied by $10^{k}$, so that all parameters can be reported to three decimal places. 

At the upper level, the estimated population tastes are statistically significant and generally consistent with established charging behavior. The positive parameter of lower-level expected utility (Log\_sum) indicates that stations providing more attractive nearby activity opportunities are more likely to be selected. Drivers prefer stations requiring shorter access distance and smaller detours and are averse to higher charging prices, as reflected by the negative parameters for Dist\_OE, Detour, and Price. \citet{visaria2022user} reported that Danish EV users were willing to make a 5.59-minute detour to save 1 Danish Krone per kWh in charging price. Using an exchange rate of 1 DKK = 0.15 US\$ , this corresponds to about $5.59/0.15=37.27$ minutes of detour per \$1/kWh reduction in charging price, which is close to our estimate of $4.882/0.144=33.90$ minutes. To compare with US studies, we convert the price parameter into an approximate out-of-pocket charging fee by assuming an average charged energy of 10.2 kWh per session \citep{evwatts2023}. The estimated population tastes imply a value of $(0.144/4.882)\times10.2=\$0.30$ per minute of detour time, or $\$18.05$ per hour. Although this estimate is substantially smaller than the $\$2.50$ per minute for reducing access time reported by \citet{ge2022charging}, the hourly value is close to \citet{usdot2025bca}'s estimates of automobile travel time (\$19.40/hour).

The coefficients of z-variable interactions reveal several meaningful patterns. For example, higher-income drivers exhibit a stronger preference for DCFC availability (+2.499) and place greater value on nearby activity opportunities (+0.245). This extends earlier evidence that income differentiates charging-preference segments \citep{wang2021electric,wen2016modeling} by identifying the specific charging attributes for which income shapes preference. Weekend sessions are associated with weaker price disutility (+0.311) and stronger preferences for DCFCs (+0.899) and nearby activity opportunities (+0.087), while longer O--D trips reduce sensitivity to access distance (+0.023) but increase sensitivity to additional detours (-0.066). These patterns further support previous findings that charging preferences vary systematically with trip purpose, day of week, and trip-chain context \citep{xu2017joint,zhang2022understanding}.

\begin{table}[!htbp]
\centering
\scriptsize
\setlength{\tabcolsep}{3pt}
\caption{Upper-level parameter estimates.}
\label{tab:upper_level_results}
\begin{tabular*}{\linewidth}{@{\extracolsep{\fill}}lrrrrrrrr@{}}
\toprule
 &  & \multicolumn{6}{c}{Standardized $z$ variables} & \\
\cmidrule(lr){3-8}
Variable & \multicolumn{1}{c}{Intercept} & \multicolumn{1}{c}{EV\_share} & \multicolumn{1}{c}{INC} & \multicolumn{1}{c}{Weekend} & \multicolumn{1}{c}{Daytime} & \multicolumn{1}{c}{Morning\_peak} & \multicolumn{1}{c}{OD\_length} & \multicolumn{1}{c@{}}{Adj.~$R^2$} \\
\midrule
Dist\_OE & $-0.111^{***}$ & $0.002^{\phantom{***}}$ & $0.009^{**\phantom{*}}$ & $0.006^{*\phantom{**}}$ & $-0.007^{**\phantom{*}}$ & $-0.020^{***}$ & $0.023^{***}$ & $0.625$ \\
 & $(0.005)^{\phantom{**}}$ & $(0.004)^{\phantom{**}}$ & $(0.004)^{\phantom{**}}$ & $(0.003)^{\phantom{**}}$ & $(0.003)^{\phantom{**}}$ & $(0.004)^{\phantom{**}}$ & $(0.002)^{\phantom{**}}$ &  \\
Detour & $-0.144^{***}$ & $-0.001^{\phantom{***}}$ & $-0.008^{\phantom{***}}$ & $0.009^{\phantom{***}}$ & $0.015^{**\phantom{*}}$ & $0.007^{\phantom{***}}$ & $-0.066^{***}$ & $0.732$ \\
 & $(0.005)^{\phantom{**}}$ & $(0.006)^{\phantom{**}}$ & $(0.006)^{\phantom{**}}$ & $(0.005)^{\phantom{**}}$ & $(0.006)^{\phantom{**}}$ & $(0.005)^{\phantom{**}}$ & $(0.008)^{\phantom{**}}$ &  \\
\addlinespace
Price & $-4.882^{***}$ & $0.461^{***}$ & $0.095^{\phantom{***}}$ & $0.311^{***}$ & $-0.294^{***}$ & $-0.485^{***}$ & $-0.065^{\phantom{***}}$ & $0.041$ \\
 & $(0.061)^{\phantom{**}}$ & $(0.071)^{\phantom{**}}$ & $(0.073)^{\phantom{**}}$ & $(0.059)^{\phantom{**}}$ & $(0.061)^{\phantom{**}}$ & $(0.063)^{\phantom{**}}$ & $(0.059)^{\phantom{**}}$ &  \\
Num\_L2 & $-0.959^{***}$ & $-1.432^{***}$ & $0.267^{\phantom{***}}$ & $-2.589^{***}$ & $-0.268^{\phantom{***}}$ & $1.213^{***}$ & $-1.324^{***}$ & $0.248$ \\
($\times 10^{2}$) & $(0.250)^{\phantom{**}}$ & $(0.277)^{\phantom{**}}$ & $(0.283)^{\phantom{**}}$ & $(0.298)^{\phantom{**}}$ & $(0.227)^{\phantom{**}}$ & $(0.222)^{\phantom{**}}$ & $(0.216)^{\phantom{**}}$ &  \\
Num\_DCFC & $9.150^{***}$ & $-1.005^{**\phantom{*}}$ & $2.499^{***}$ & $0.899^{***}$ & $-2.553^{***}$ & $-1.924^{***}$ & $-0.980^{***}$ & $0.100$ \\
($\times 10^{2}$) & $(0.377)^{\phantom{**}}$ & $(0.412)^{\phantom{**}}$ & $(0.437)^{\phantom{**}}$ & $(0.332)^{\phantom{**}}$ & $(0.331)^{\phantom{**}}$ & $(0.377)^{\phantom{**}}$ & $(0.297)^{\phantom{**}}$ &  \\
\addlinespace
ChargePoint & $0.486^{***}$ & $0.150^{\phantom{***}}$ & $-0.736^{***}$ & $-0.194^{**\phantom{*}}$ & $-0.221^{*\phantom{**}}$ & $0.033^{\phantom{***}}$ & $0.059^{\phantom{***}}$ & $0.311$ \\
 & $(0.106)^{\phantom{**}}$ & $(0.108)^{\phantom{**}}$ & $(0.123)^{\phantom{**}}$ & $(0.098)^{\phantom{**}}$ & $(0.127)^{\phantom{**}}$ & $(0.099)^{\phantom{**}}$ & $(0.088)^{\phantom{**}}$ &  \\
Tesla & $-1.529^{***}$ & $0.164^{\phantom{***}}$ & $-0.428^{***}$ & $0.109^{\phantom{***}}$ & $-0.349^{**\phantom{*}}$ & $-0.277^{**\phantom{*}}$ & $0.169^{\phantom{***}}$ & $0.173$ \\
 & $(0.129)^{\phantom{**}}$ & $(0.136)^{\phantom{**}}$ & $(0.150)^{\phantom{**}}$ & $(0.116)^{\phantom{**}}$ & $(0.143)^{\phantom{**}}$ & $(0.125)^{\phantom{**}}$ & $(0.104)^{\phantom{**}}$ &  \\
Other & $0.593^{***}$ & $0.176^{*\phantom{**}}$ & $-0.688^{***}$ & $0.034^{\phantom{***}}$ & $-0.190^{\phantom{***}}$ & $-0.209^{**\phantom{*}}$ & $-0.064^{\phantom{***}}$ & $0.261$ \\
 & $(0.104)^{\phantom{**}}$ & $(0.105)^{\phantom{**}}$ & $(0.118)^{\phantom{**}}$ & $(0.094)^{\phantom{**}}$ & $(0.125)^{\phantom{**}}$ & $(0.098)^{\phantom{**}}$ & $(0.087)^{\phantom{**}}$ &  \\
\addlinespace
Log\_sum & $1.428^{***}$ & $-0.189^{***}$ & $0.245^{***}$ & $0.087^{*\phantom{**}}$ & $0.013^{\phantom{***}}$ & $-0.080^{**\phantom{*}}$ & $-0.118^{***}$ & $0.022$ \\
 & $(0.043)^{\phantom{**}}$ & $(0.048)^{\phantom{**}}$ & $(0.045)^{\phantom{**}}$ & $(0.049)^{\phantom{**}}$ & $(0.042)^{\phantom{**}}$ & $(0.039)^{\phantom{**}}$ & $(0.032)^{\phantom{**}}$ &  \\
\bottomrule
\end{tabular*}
\par\smallskip
\begin{minipage}{\linewidth}
\textit{Notes:} Intercept reflects the population-level tastes; each $z$ column is the change in the taste for a one-standard-deviation increase in that variable. $^{***}p<0.01$, $^{**}p<0.05$, $^{*}p<0.1$.
\end{minipage}
\end{table}

At the lower level, the population tastes also exhibit behaviorally consistent patterns. The parameters for all eight POI categories are positive and statistically significant, indicating a similar finding that a greater supply of nearby activities increases the probability of visiting a POI during charging \citep{visaria2022user,ten2020investigating}. The visual score parameters further show that drivers are more attracted to environments perceived as beautiful, lively, safe, and wealthy, while boring and depressing environments reduce the utility. These visual-environment effects extend the existing charging literature by quantifying the effects of perceived environmental quality. 

The estimated z effects also reveal intuitive behavioral insights. Higher-income drivers place greater weight on restaurants (+0.058), grocery stores (+0.394), entertainment (+0.483), and environments perceived as lively (+0.233), safe (+0.414), and wealthy (+0.090). Morning-peak sessions, in contrast, reduce the valuation of restaurants (-0.804), retail (-0.159), beauty stores (-0.789), and wealthy-looking environments (-0.697), while increasing the valuation of errand POIs (+0.880). Such temporal shifts are consistent with previous findings that public-charging behavior varies by the timing of charging \citep{ma2022research,zhang2022understanding}, while our study further quantifies how these temporal contexts systematically alter preferences for specific attributes.

\begin{table}[!htbp]
\centering
\scriptsize
\setlength{\tabcolsep}{3pt}
\caption{Lower-level parameter estimates.}
\label{tab:lower_level_results}
\begin{tabular*}{\linewidth}{@{\extracolsep{\fill}}lrrrrrrrr@{}}
\toprule
 &  & \multicolumn{6}{c}{Standardized $z$ variables} & \\
\cmidrule(lr){3-8}
Variable & \multicolumn{1}{c}{Intercept} & \multicolumn{1}{c}{EV\_share} & \multicolumn{1}{c}{INC} & \multicolumn{1}{c}{Weekend} & \multicolumn{1}{c}{Daytime} & \multicolumn{1}{c}{Morning\_peak} & \multicolumn{1}{c}{OD\_length} & \multicolumn{1}{c@{}}{Adj.~$R^2$} \\
\midrule
\multicolumn{9}{@{}l}{\textit{POIs within 200 m}} \\
Fitness & $2.658^{***}$ & $0.164^{\phantom{***}}$ & $0.093^{\phantom{***}}$ & $0.391^{\phantom{***}}$ & $0.274^{\phantom{***}}$ & $-0.523^{\phantom{***}}$ & $0.270^{\phantom{***}}$ & $0.536$ \\
($\times 10^{3}$) & $(0.909)^{\phantom{**}}$ & $(0.483)^{\phantom{**}}$ & $(0.345)^{\phantom{**}}$ & $(1.346)^{\phantom{**}}$ & $(0.189)^{\phantom{**}}$ & $(0.331)^{\phantom{**}}$ & $(0.310)^{\phantom{**}}$ &  \\
Pharmacies & $1.729^{***}$ & $0.537^{*\phantom{**}}$ & $0.268^{\phantom{***}}$ & $-0.010^{\phantom{***}}$ & $0.229^{\phantom{***}}$ & $-0.069^{\phantom{***}}$ & $-0.073^{\phantom{***}}$ & $0.894$ \\
($\times 10^{2}$) & $(0.411)^{\phantom{**}}$ & $(0.319)^{\phantom{**}}$ & $(0.220)^{\phantom{**}}$ & $(0.158)^{\phantom{**}}$ & $(0.248)^{\phantom{**}}$ & $(0.196)^{\phantom{**}}$ & $(0.065)^{\phantom{**}}$ &  \\
Restaurants & $1.306^{**\phantom{*}}$ & $-0.015^{**\phantom{*}}$ & $0.058^{***}$ & $0.072^{\phantom{***}}$ & $0.032^{***}$ & $-0.804^{**\phantom{*}}$ & $0.022^{\phantom{***}}$ & $0.893$ \\
($\times 10^{3}$) & $(0.521)^{\phantom{**}}$ & $(0.007)^{\phantom{**}}$ & $(0.010)^{\phantom{**}}$ & $(0.073)^{\phantom{**}}$ & $(0.005)^{\phantom{**}}$ & $(0.388)^{\phantom{**}}$ & $(0.014)^{\phantom{**}}$ &  \\
Retail & $1.012^{***}$ & $-0.008^{\phantom{***}}$ & $0.012^{\phantom{***}}$ & $0.254^{***}$ & $0.113^{***}$ & $-0.159^{***}$ & $-0.104^{***}$ & $0.260$ \\
($\times 10^{4}$) & $(0.026)^{\phantom{**}}$ & $(0.009)^{\phantom{**}}$ & $(0.012)^{\phantom{**}}$ & $(0.044)^{\phantom{**}}$ & $(0.007)^{\phantom{**}}$ & $(0.010)^{\phantom{**}}$ & $(0.009)^{\phantom{**}}$ &  \\
Grocery & $1.444^{***}$ & $-0.070^{\phantom{***}}$ & $0.394^{***}$ & $0.194^{\phantom{***}}$ & $0.195^{***}$ & $-0.104^{\phantom{***}}$ & $-0.032^{\phantom{***}}$ & $0.588$ \\
($\times 10^{3}$) & $(0.176)^{\phantom{**}}$ & $(0.149)^{\phantom{**}}$ & $(0.120)^{\phantom{**}}$ & $(0.167)^{\phantom{**}}$ & $(0.072)^{\phantom{**}}$ & $(0.092)^{\phantom{**}}$ & $(0.113)^{\phantom{**}}$ &  \\
Beauty & $2.626^{***}$ & $0.034^{\phantom{***}}$ & $0.165^{**\phantom{*}}$ & $0.856^{*\phantom{**}}$ & $0.177^{***}$ & $-0.789^{***}$ & $-0.267^{***}$ & $0.629$ \\
($\times 10^{4}$) & $(0.311)^{\phantom{**}}$ & $(0.065)^{\phantom{**}}$ & $(0.069)^{\phantom{**}}$ & $(0.503)^{\phantom{**}}$ & $(0.036)^{\phantom{**}}$ & $(0.143)^{\phantom{**}}$ & $(0.034)^{\phantom{**}}$ &  \\
Errands & $3.374^{***}$ & $-0.127^{\phantom{***}}$ & $0.087^{\phantom{***}}$ & $-1.289^{***}$ & $0.056^{\phantom{***}}$ & $0.880^{**\phantom{*}}$ & $-0.022^{\phantom{***}}$ & $0.879$ \\
($\times 10^{2}$) & $(0.411)^{\phantom{**}}$ & $(0.113)^{\phantom{**}}$ & $(0.062)^{\phantom{**}}$ & $(0.276)^{\phantom{**}}$ & $(0.044)^{\phantom{**}}$ & $(0.424)^{\phantom{**}}$ & $(0.029)^{\phantom{**}}$ &  \\
Entertainment & $1.757^{***}$ & $-0.030^{\phantom{***}}$ & $0.483^{***}$ & $-0.061^{\phantom{***}}$ & $0.262^{***}$ & $-0.041^{\phantom{***}}$ & $-0.056^{\phantom{***}}$ & $0.589$ \\
($\times 10^{3}$) & $(0.218)^{\phantom{**}}$ & $(0.158)^{\phantom{**}}$ & $(0.132)^{\phantom{**}}$ & $(0.102)^{\phantom{**}}$ & $(0.068)^{\phantom{**}}$ & $(0.104)^{\phantom{**}}$ & $(0.106)^{\phantom{**}}$ &  \\
\addlinespace
\multicolumn{9}{@{}l}{\textit{Perceived streetscape}} \\
Beautiful & $1.045^{***}$ & $-0.015^{\phantom{***}}$ & $0.033^{***}$ & $0.028^{**\phantom{*}}$ & $0.057^{***}$ & $0.133^{***}$ & $0.002^{\phantom{***}}$ & $0.017$ \\
($\times 10^{4}$) & $(0.011)^{\phantom{**}}$ & $(0.013)^{\phantom{**}}$ & $(0.013)^{\phantom{**}}$ & $(0.012)^{\phantom{**}}$ & $(0.010)^{\phantom{**}}$ & $(0.011)^{\phantom{**}}$ & $(0.011)^{\phantom{**}}$ &  \\
Boring & $-3.904^{***}$ & $-0.266^{\phantom{***}}$ & $-0.539^{\phantom{***}}$ & $1.034^{***}$ & $0.047^{\phantom{***}}$ & $-0.374^{\phantom{***}}$ & $0.081^{\phantom{***}}$ & $0.913$ \\
($\times 10^{2}$) & $(0.363)^{\phantom{**}}$ & $(0.433)^{\phantom{**}}$ & $(0.365)^{\phantom{**}}$ & $(0.299)^{\phantom{**}}$ & $(0.322)^{\phantom{**}}$ & $(0.366)^{\phantom{**}}$ & $(0.196)^{\phantom{**}}$ &  \\
Depressing & $-3.578^{***}$ & $-0.082^{***}$ & $-0.128^{***}$ & $-0.063^{**\phantom{*}}$ & $0.067^{***}$ & $0.327^{***}$ & $0.009^{\phantom{***}}$ & $0.099$ \\
($\times 10^{4}$) & $(0.027)^{\phantom{**}}$ & $(0.029)^{\phantom{**}}$ & $(0.028)^{\phantom{**}}$ & $(0.027)^{\phantom{**}}$ & $(0.020)^{\phantom{**}}$ & $(0.023)^{\phantom{**}}$ & $(0.021)^{\phantom{**}}$ &  \\
Lively & $2.948^{***}$ & $-0.901^{***}$ & $0.223^{***}$ & $0.094^{\phantom{***}}$ & $-1.156^{***}$ & $-0.077^{*\phantom{**}}$ & $-0.101^{***}$ & $0.871$ \\
($\times 10^{2}$) & $(0.410)^{\phantom{**}}$ & $(0.160)^{\phantom{**}}$ & $(0.049)^{\phantom{**}}$ & $(0.063)^{\phantom{**}}$ & $(0.409)^{\phantom{**}}$ & $(0.043)^{\phantom{**}}$ & $(0.020)^{\phantom{**}}$ &  \\
Safety & $2.792^{***}$ & $-0.263^{***}$ & $0.414^{***}$ & $-0.012^{\phantom{***}}$ & $0.187^{***}$ & $0.865^{***}$ & $-0.080^{***}$ & $0.216$ \\
($\times 10^{4}$) & $(0.083)^{\phantom{**}}$ & $(0.040)^{\phantom{**}}$ & $(0.036)^{\phantom{**}}$ & $(0.034)^{\phantom{**}}$ & $(0.032)^{\phantom{**}}$ & $(0.103)^{\phantom{**}}$ & $(0.027)^{\phantom{**}}$ &  \\
Wealthy & $3.173^{***}$ & $0.090^{**\phantom{*}}$ & $0.429^{***}$ & $-0.349^{***}$ & $0.084^{***}$ & $-0.697^{***}$ & $0.495^{***}$ & $0.363$ \\
($\times 10^{4}$) & $(0.233)^{\phantom{**}}$ & $(0.041)^{\phantom{**}}$ & $(0.043)^{\phantom{**}}$ & $(0.039)^{\phantom{**}}$ & $(0.028)^{\phantom{**}}$ & $(0.135)^{\phantom{**}}$ & $(0.098)^{\phantom{**}}$ &  \\
\addlinespace
\multicolumn{9}{@{}l}{\textit{Available time and constant}} \\
Stay\_dur & $1.971^{***}$ & $0.034^{\phantom{***}}$ & $0.116^{\phantom{***}}$ & $0.206^{\phantom{***}}$ & $0.136^{\phantom{***}}$ & $0.680^{***}$ & $-0.165^{***}$ & $0.889$ \\
($\times 10^{3}$) & $(0.152)^{\phantom{**}}$ & $(0.178)^{\phantom{**}}$ & $(0.170)^{\phantom{**}}$ & $(0.145)^{\phantom{**}}$ & $(0.138)^{\phantom{**}}$ & $(0.148)^{\phantom{**}}$ & $(0.054)^{\phantom{**}}$ &  \\
ASC\_visit & $-2.276^{***}$ & $0.376^{**\phantom{*}}$ & $0.183^{\phantom{***}}$ & $-0.219^{*\phantom{**}}$ & $0.437^{**\phantom{*}}$ & $-0.014^{\phantom{***}}$ & $0.058^{\phantom{***}}$ & $0.902$ \\
 & $(0.193)^{\phantom{**}}$ & $(0.169)^{\phantom{**}}$ & $(0.139)^{\phantom{**}}$ & $(0.113)^{\phantom{**}}$ & $(0.197)^{\phantom{**}}$ & $(0.137)^{\phantom{**}}$ & $(0.072)^{\phantom{**}}$ &  \\
\bottomrule
\end{tabular*}
\par\smallskip
\begin{minipage}{\linewidth}
\textit{Notes:} Intercept reflects the population-level tastes; each $z$ column is the change in the taste for a one-standard-deviation increase in that variable. $^{***}p<0.01$, $^{**}p<0.05$, $^{*}p<0.1$.
\end{minipage}
\end{table}

\subsubsection{Empirical taste distribution}
\label{sec:result_taste_distribution}
\begin{figure}[htbp]
  \centering
  \includegraphics[width=0.75\linewidth]{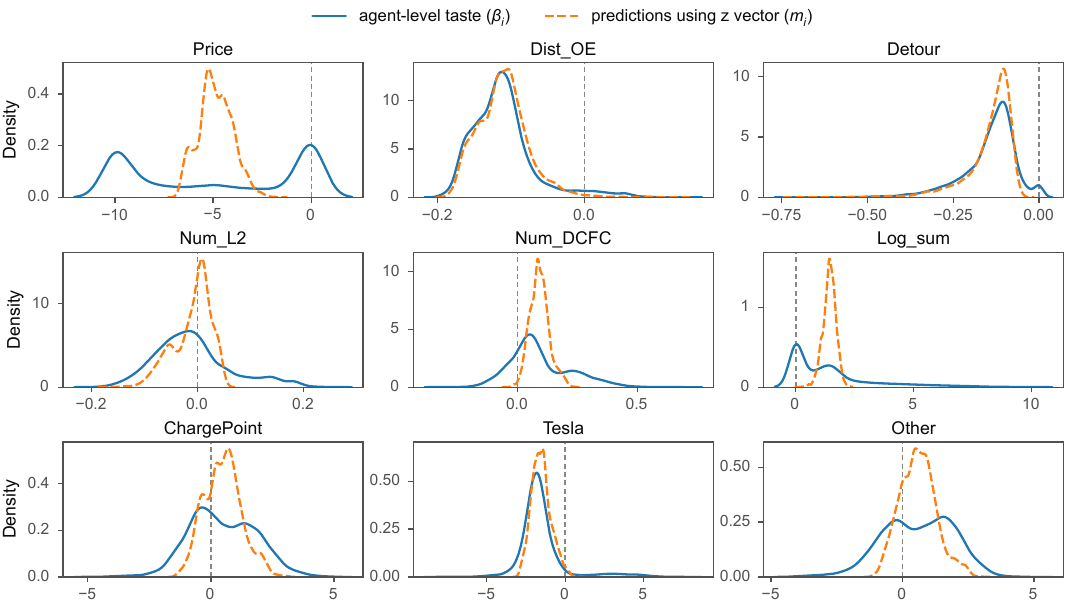}
  \caption{Parameter distribution (upper-level).}
  \label{fig:taste_distribution_upper}
\end{figure}

\begin{figure}[htbp]
  \centering
  \includegraphics[width=\linewidth]{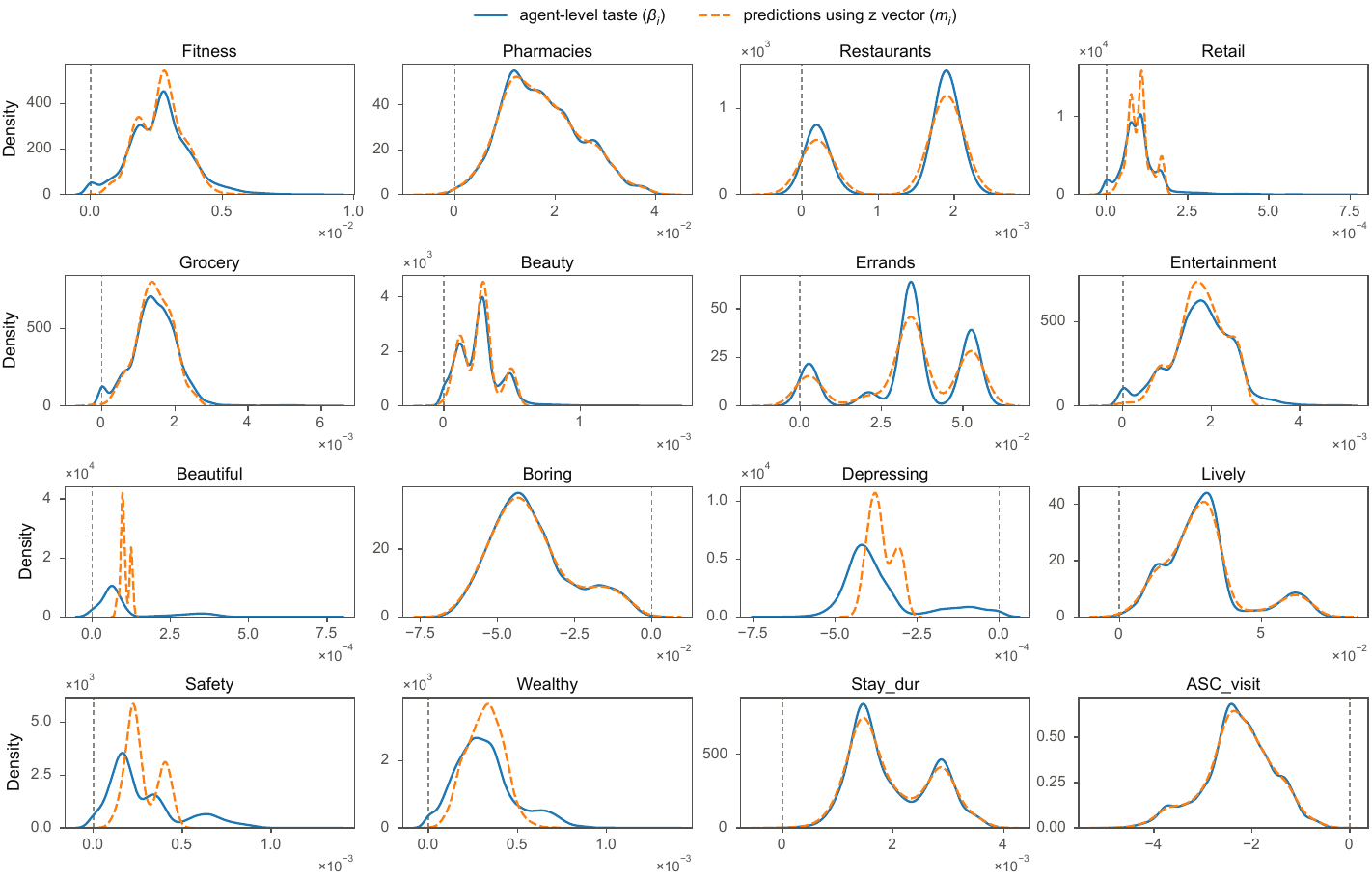}
  \caption{Parameter distribution (lower-level).}
  \label{fig:taste_distribution_lower}
\end{figure}

The $\mathrm{Adj.}R^2$ values reported in \cref{tab:upper_level_results} and \cref{tab:lower_level_results} provide a first indication of how much of the estimated taste heterogeneity can be explained by the z variables. \cref{fig:taste_distribution_upper} and \cref{fig:taste_distribution_lower} provide a more direct view of this relationship by comparing the empirical distributions of the estimated agent-level tastes ($\vect{\beta}_i$), with their corresponding z-based predictions ($\vect{m}_i$). The dispersion of the blue distributions reflects the magnitude of taste heterogeneity across charging sessions, while the degree of overlap between the blue and orange distributions indicates how much of this heterogeneity can be explained by the z variables. Larger discrepancies between the two distributions therefore suggest a greater role for residual agent-level heterogeneity, which may reflect unobserved preferences as well as noise in the GPS-derived contextual variables.

At the upper level (\cref{fig:taste_distribution_upper}), preferences for access distance and detour time are relatively concentrated, suggesting that drivers respond to these spatial costs in a comparatively homogeneous way. In contrast, charging price exhibits a much more dispersed and multimodal empirical distribution, indicating substantial heterogeneity in price sensitivity across charging sessions. Considerable dispersion is also observed for the number of Level 2 and DCFC ports and for the log-sum utility of nearby activity opportunities. The comparison between empirical and z-predicted tastes further reveals that the observed context explains some attributes much better than others. The distributions for distance and detour time largely overlap. By contrast, the predicted distributions for price, charger counts, and log-sum utility are much more concentrated than the corresponding empirical distributions. This indicates that a substantial share of heterogeneity in these preferences remains unexplained by the z vector and must instead be recovered through the agent-level deviation component of AMXL-z.

A similar pattern emerges at the lower level (\cref{fig:taste_distribution_lower}), although the explanatory power of z variables is generally stronger. The empirical taste distributions for pharmacies, grocery stores, entertainment, boring and lively streetscapes, stay duration, and the visit constant are relatively concentrated and well-recovered by the z-based predictions. Other tastes exhibit substantially greater residual variation. For example, beautiful, depressing, safety, and wealthy visual perceptions show noticeable discrepancies between the empirical and predicted distributions. Taken together, these results show that some tastes are more homogeneous and can be largely reconstructed from the available z variables, whereas others contain substantial residual heterogeneity that would be missed by a model relying exclusively on observed characteristics.

\subsection{Policy recommendations for PEVCS deployment}
\label{sec:policy_recommendation}
As EV adoption continues to grow, PEVCS deployment is expected to expand accordingly. The results from our model provide several implications for future PEVCS planning and related behavioral policies.

First, public charging availability should be evaluated from an activity-chain perspective rather than simply by station density or proximity. The time budget defined by the activities before (O) and after (D) charging determines which stations are realistically available to a driver. In our data, an observed charging session has about 85 alternatives on average, far more than the small choice sets typically assumed in SP surveys. The same time budget also determines how much time remains for charging and nearby activities. PEVCS planning should therefore account for trip-chain constraints and the temporal accessibility of stations.

Second, PEVCS deployment should be coordinated with the surrounding built environment. The significant positive log-sum parameter shows that the expected utility of nearby activities contributes directly to charging-station choice. At the lower level, both POI proximity and visual perceptions significantly affect activity utility. These results suggest that a charging station should not be planned as stand-alone infrastructure. Co-locating chargers with useful activity opportunities and improving the surrounding streetscape environment can increase the attractiveness of both the charging location and the activities undertaken during charging. This also creates an opportunity to integrate PEVCS deployment with local commercial and community development.

Third, behavioral interventions should explicitly account for taste heterogeneity. Some preferences are relatively homogeneous and can be explained well by observed driver and trip contexts (z variable). For example, preferences for access distance and detour time are comparatively concentrated, with $\mathrm{Adj.}R^2$ values of 0.625 and 0.732. These attributes provide relatively stable policy levers. Other preferences, such as charging price, charger counts, and nearby activity opportunities, exhibit much greater residual heterogeneity and are weakly explained by the available z variables. Uniform incentives based on these attributes may therefore produce uneven responses. Policies targeting such preferences should be accompanied by better measurement of driver and trip characteristics and further investigation of the agent-level deviations that remain unexplained.

\section{Conclusion}
\label{sec:conclusion}
This study develops a framework for modeling public EV charging as a joint decision involving both charging-station selection and nearby activity participation. We construct session-specific charging choice sets from large-scale GPS trajectories and develop a nested choice structure. We further extend the AMXL to AMXL-z, which separates taste heterogeneity associated with observed driver and charging-context characteristics from residual heterogeneity learned directly from data.

The study makes two main contributions. First, it provides a RP framework for jointly modeling charging station choice and nearby POI visitation from GPS data. This moves beyond conventional charging models that treat charging as an isolated decision or rely on SP surveys. Second, AMXL-z provides a flexible approach to model taste heterogeneity for large-scale GPS data. By decomposing agent-level tastes into a population component, systematic variation explained by the z variables, and empirically learned residual deviations, the model leverages the rich observations available in GPS data while remaining robust to the limited and noisy information.

The simulation study demonstrates the value of this decomposition under different levels of data noise and panel information. AMXL-quadratic generally provides the strongest predictive performance and parameter recovery, while the simpler AMXL-linear specification becomes increasingly competitive as noise increases. The simulations also highlight their computational advantages for large mobility datasets.

The San Francisco Bay Area case study further shows that public charging behavior is jointly shaped by charging infrastructure, activity opportunities, and travel context. Drivers prefer shorter access distances and detours and lower charging prices, while the significant positive log-sum parameter confirms that attractive nearby activities also increase usage. The empirical taste distributions reveal that these preferences are not equally heterogeneous. Distance and detour preferences are relatively concentrated and can be explained reasonably well by observed context, whereas preferences for price, charger availability, and nearby activity opportunities contain substantially more residual variation. These findings provide valuable behavioral insights for future PEVCS deployment.

Several extensions remain for future research. First, the current AMXL-z application considers linear and quadratic mappings between z variables and taste parameters. More flexible functional forms could be incorporated, such as neural networks. Second, the framework is not specific to EV charging and can be extended to other large-scale GPS-based choice settings in which observed characteristics explain only part of behavioral heterogeneity. Finally, the current lower-level model represents nearby activity participation as a binary visit-or-not decision. Future work could model the choice among activity categories to provide a richer representation of how charging is embedded within daily activity patterns.


\ifdoubleblind
\else
\section*{CRediT authorship contribution statement}
\textbf{Xiyuan (Ryan) Ren:} Conceptualization, Methodology, Software, Formal analysis, Visualization, Writing. \textbf{Callie Clark:} Conceptualization, Data curation, Formal analysis, Visualization, Writing. \textbf{Yunyi (Lauren) Wang:} Software, Formal analysis, Writing. \textbf{Takahiro Yabe:} Conceptualization, Supervision, Writing, Funding acquisition. \textbf{Joseph Y.J. Chow:} Conceptualization, Supervision, Methodology, Funding acquisition, Writing.

\section*{Acknowledgements}
Takahiro Yabe and Joseph Y. J. Chow acknowledge support by the National Science Foundation under grant number BCS-2425021.

\section*{Data and code availability}
The replication package for the simulation study is available at
\url{https://github.com/xr2006/AMXL-z-simulation-study}. 
The mobility data that support the findings of this study are available from Cuebiq through their Social Impact program, but restrictions apply to the availability of these data, which were used under the license for the current study and are therefore not publicly available. Information about how to request access to the data and its conditions and limitations can be found in \url{https://cuebiq.com/social-impact/}.  
Data access requests should be submitted through Cuebiq's Social Impact customer page \url{https://cuebiq.com/demo/}, where the Sales team at Cuebiq may be contacted in a timely manner. 

\section*{Declaration of competing interest}
The authors declare no competing interests.

\section*{Declaration of generative AI and AI-assisted technologies in the writing process}
During the preparation of this work the authors used ChatGPT 5.6 and Claude Opus 5.5 in order to improve readability and language. After using this tool/service, the authors reviewed and edited the content as needed and take full responsibility for the content of the published article.

\appendix
\section{}
\label{app:main}

\subsection{Data-generating process}
\label{app:dgp}
This appendix describes how the synthetic datasets are generated and documents the distributions of the true taste parameters.

\textit{Driver characteristics}. We draw 2000 drivers independently. Their characteristics follow the distributions listed in \cref{tab:synthetic-z-vector}. Annual income (INCOME), measured in thousands of dollars, is log-normal with parameters $\ln{60}$ and 0.35, so that the median income is \$60,000. The dummy time-of-day variable TOD equals one if the driver usually charges at night and is Bernoulli distributed with probability 0.5, and the average charging duration, DUR, is 30, 60 or 90 minutes with equal probabilities. The three characteristics are independent. They enter the tastes in standardized form, $\vect{z_n}=(inc_n,tod_n,dur_n)$, with $inc_n=(\ln Income_n -\ln 60)/0.35$, $tod_n=TOD_n$, and $dur_n=(DUR_n-60)/30$.

\textit{Station attributes}. For each choice observation, the attributes of the available alternatives are drawn independently. Travel time to the station (TT) follows a uniform distribution between 0 and 10 minutes, the number of nearby points of interest (POI) is drawn uniformly between 0 and 20, and charging cost (CO) follows a uniform distribution between \$0 and \$6. These attributes vary across both alternatives and choice observations and enter the utility function directly.

\begin{table}[htbp]
  \centering
  \begin{threeparttable}
    \caption{Description of driver characteristics}
    \label{tab:synthetic-z-vector}
    \footnotesize
    
    \begin{tabular}{@{} L{1.25in} l L{1.60in} L{1.60in} l @{}}
      \toprule
      Level & Variable & Description & Distribution & Unit \\
      \midrule
      \multirow{3}{1.25in}{Driver characteristics $z_n$}
        & INCOME & Annual income
          & LogNormal($\ln 60$, $0.35^2$); median 60 & \$1000 \\
      \addlinespace
        & TOD & Usually charges at night ($1=$ night, $0=$ day)
          & Bernoulli(0.5) & -- \\
      \addlinespace
        & DUR & Average charging duration
          & Uniform on $\{30, 60, 90\}$ & min \\
      \addlinespace
      \multirow{3}{1.25in}{Station attributes, drawn on every choice observation}
        & TT & Travel time to the station & Uniform(0, 10) & min \\
      \addlinespace
        & POI & Number of points of interest near the station
          & Uniform(0, 20) & count \\
      \addlinespace
        & CO & Charging cost & Uniform(0, 6) & \$ \\
      \bottomrule
    \end{tabular}
    \begin{tablenotes}[flushleft]
      \footnotesize
      \item \textit{Notes:} $z_n$ denotes the characteristics of driver $n$.
        Driver-level variables are drawn once per driver; station-level
        variables are redrawn on every choice observation.
    \end{tablenotes}
  \end{threeparttable}
\end{table}

\textit{Taste functions}. \cref{fig:dgp_taste_function} shows the systematic taste functions. These functions are designed to mimic real charging preferences. For example, travel time is generally associated with negative utility, with its disutility increasing with income and being weakened by longer charging durations. During daytime, POI utility increases with both income and charging duration, whereas at night the positive effect is concentrated among longer charging sessions and becomes weak or even negative for short-duration charging.

\begin{figure}[htbp]
  \centering
  \includegraphics[width=0.8\linewidth]{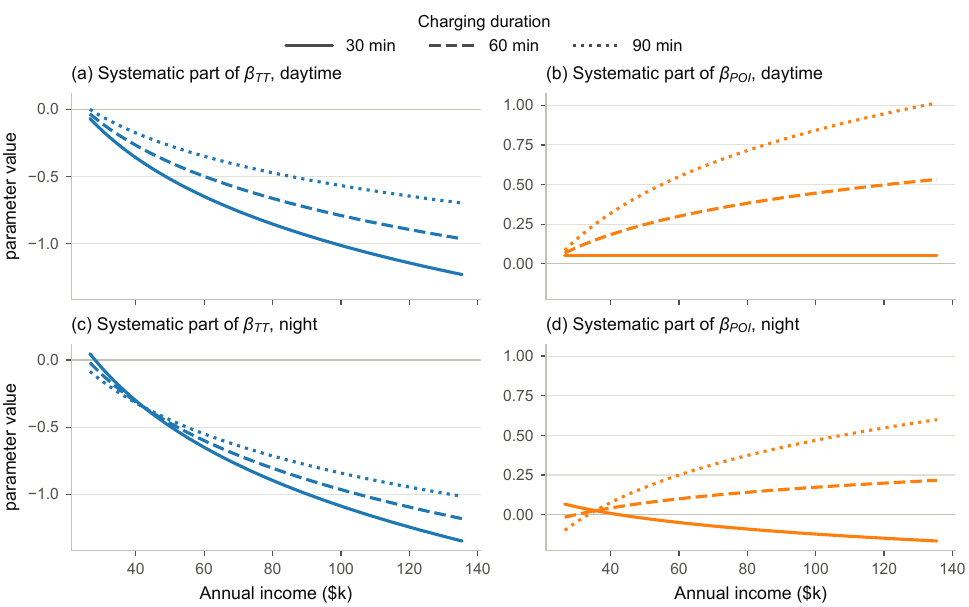}
  \caption{Systematic tastes against annual income, by time of day and charging duration, over the 1st–99th percentiles of income.}
  \label{fig:dgp_taste_function}
\end{figure}

\textit{Random components}. We generate the random part of taste as \cref{eq:residual_generation}. 
\begin{equation}
\eta_{k,n} = \sigma s_k(\vect{z}_n)u_{k,n}, \qquad k \in \{TT,POI\}
\label{eq:residual_generation}
\end{equation}
where the random component is governed by three components. First, $\sigma$ controls the overall level of randomness, referring to the noise level. Second, $s_k(\vect{z}_n)=\sum_{m}^{P} c_{k,m}^{2}\phi_m(\vect{z})^2$ links the magnitude of the variation to the $P=7$ characteristics (three driver characteristics, three interaction terms, and one intercept), so that the same noise can generate different amounts of residual taste variation across drivers with different $\vect{z}$. This is intended to mimic GPS-based datasets, in which missing or spatially aggregated individual characteristics may introduce heterogeneous measurement uncertainty across individuals. Third, $u_{k,n}$ is a standardized random component that determines the shape of the residual distribution. We define $u_{TT,n}=1-G_{TT,n}$, $u_{POI,n}=G_{POI,n}-1$, and $G_{k,n}\sim\operatorname{Exponential}(1)$ such that both $u_{TT,n}$ and $u_{POI,n}$ have mean zero and variance one, but exhibit negative and positive skewness, respectively. \cref{fig:dgp_taste_distribution} shows the true taste of all 2000 drivers against annual income.

\begin{figure}[htbp]
  \centering
  \includegraphics[width=0.8\linewidth]{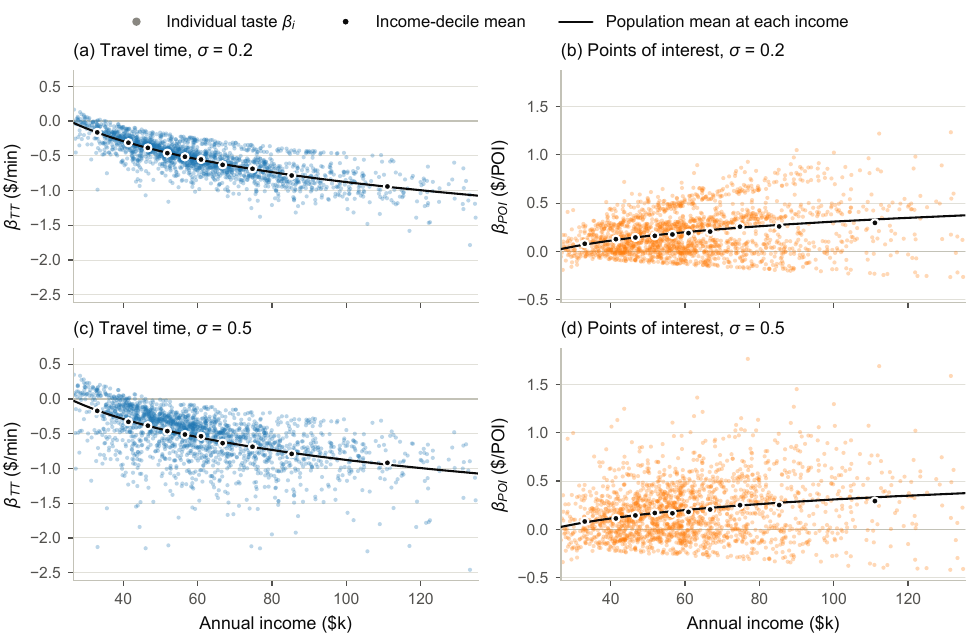}
  \caption{True tastes of all 2000 drivers against annual income. Points: individual tastes; black markers: means within income deciles; black line: population mean taste.}
  \label{fig:dgp_taste_distribution}
\end{figure}

\begin{table}[htbp]
  \centering
  \begin{threeparttable}
    \caption{Summary of true individual-level tastes and the random component across data splits}
    \label{tab:train_dev_test_split}
    \small
    \newcolumntype{M}{S[table-format=-1.3]}   
    \newcolumntype{D}{S[table-format=1.3]}    
    \newcommand{\ci}[2]{$[#1,\; #2]$}
    \begin{tabular*}{\linewidth}{@{\extracolsep{\fill}} l M D c M D c @{}}
      \toprule
      & \multicolumn{3}{c}{$\beta_{TT,n}$ (\$/min)}
      & \multicolumn{3}{c}{$\beta_{POI,n}$ (\$/POI)} \\
      \cmidrule(lr){2-4} \cmidrule(l){5-7}
      Sample (drivers) & {Mean} & {SD} & 95\% CI of $\eta$
                       & {Mean} & {SD} & 95\% CI of $\eta$ \\
      \midrule
      \multicolumn{7}{@{}l}{$\sigma = 0.2$} \\
      Training (1200)   & -0.536 & 0.273 & \ci{-0.292}{0.115}
                        &  0.185 & 0.229 & \ci{-0.089}{0.197} \\
      Development (400) & -0.562 & 0.291 & \ci{-0.308}{0.117}
                        &  0.195 & 0.234 & \ci{-0.085}{0.235} \\
      Test (400)        & -0.545 & 0.258 & \ci{-0.295}{0.108}
                        &  0.201 & 0.238 & \ci{-0.087}{0.222} \\
      All (2000)        & -0.543 & 0.274 & \ci{-0.295}{0.115}
                        &  0.190 & 0.232 & \ci{-0.088}{0.214} \\
      \midrule
      \multicolumn{7}{@{}l}{$\sigma = 0.5$} \\
      Training (1200)   & -0.536 & 0.377 & \ci{-0.730}{0.288}
                        &  0.177 & 0.292 & \ci{-0.223}{0.492} \\
      Development (400) & -0.556 & 0.385 & \ci{-0.771}{0.293}
                        &  0.198 & 0.300 & \ci{-0.213}{0.589} \\
      Test (400)        & -0.551 & 0.353 & \ci{-0.738}{0.270}
                        &  0.199 & 0.302 & \ci{-0.218}{0.555} \\
      All (2000)        & -0.543 & 0.374 & \ci{-0.738}{0.287}
                        &  0.185 & 0.295 & \ci{-0.221}{0.534} \\
      \bottomrule
    \end{tabular*}
    \begin{tablenotes}[flushleft]
      \footnotesize
      \item \textit{Notes:} Mean and SD are reported for the true individual-level taste parameters $\beta_{k,n}$, while the 95\% confidence intervals are reported for the residual component $\eta_{k,n}$.
    \end{tablenotes}
  \end{threeparttable}
\end{table}

We use these asymmetric residual distributions rather than the Gaussian mixing distribution commonly assumed in MXL to reflect the possibility that preference variation left unexplained by incomplete or aggregated GPS-derived characteristics is heterogeneous and asymmetric. \cref{tab:train_dev_test_split} summarizes the distribution of the true individual-level taste parameters across the training, development, test, and full samples under the two noise levels. For each taste parameter, the table reports the mean and standard deviation of $\beta_{k,n}$, together with the 95\% confidence interval of the corresponding random component $\eta_{k,n}$. The three data splits exhibit similar taste distributions, indicating that the random partition preserves the overall heterogeneity of the synthetic population. Increasing the noise level from $\sigma=0.2$ to $\sigma=0.5$ leaves the mean tastes largely unchanged but increases their dispersion.

\subsection{Examples of street-view panoramas and visual perception scores}
\cref{fig:gsv_examples} presents three representative PEVCSs with distinct surrounding environments. For each station, the 360$^\circ$ panorama is retrieved from GSV at the station's geolocation. Following \citet{yao2019human}, each panorama is first semantically segmented into visual elements such as sky, buildings, vegetation, roads and vehicles, and the resulting visual features are then used to score six perception dimensions on a 0--100 scale.

The three examples show how these scores vary with the streetscape. EVCS\_548 is located next to modern office buildings and a landscaped plaza with outdoor seating, and scores highest on wealthy (72) and lively (64). EVCS\_539 is surrounded by mature trees and low-rise buildings, and scores highest on beautiful (69). EVCS\_91 faces a wide arterial road and a parking structure under construction, and scores highest on depressing (73) and boring (67). These differences in perceived streetscape quality are incorporated into the attributes for the lower-level POI visitation utility.

\begin{figure}[htbp]
    \centering
    \includegraphics[width=\textwidth]{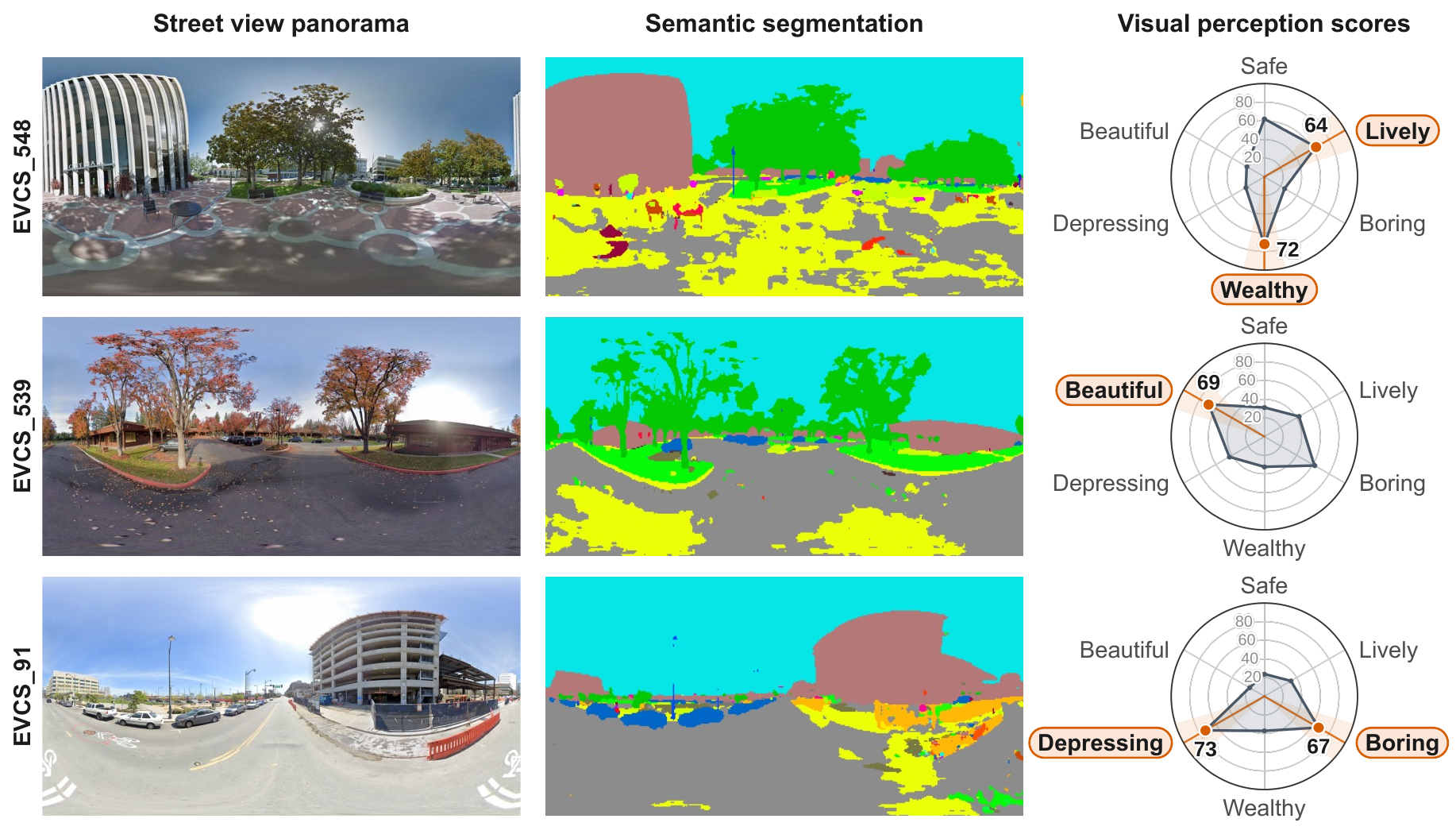}
    \caption{Street-view panoramas, semantic segmentation results and visual perception scores for three representative PEVCSs.}
    \label{fig:gsv_examples}
\end{figure}

\subsection{Variables descriptions}
\label{app:variable_descriptions}
\cref{tab:variable_definitions} summarizes the name, symbol, data source, and description of varibles used in our model.

\begingroup
\footnotesize
\setlength{\LTleft}{0pt}
\setlength{\LTright}{\fill}
\begin{xltabular}{\linewidth}{@{}L{1.05in} l L{1.05in} Y@{}}
\caption{Variables used in the case study.}
\label{tab:variable_definitions} \\
\toprule
Variable name & Symbol & Data source & Description \\
\midrule
\endfirsthead
\caption[]{\textit{continued}} \\
\toprule
Variable name & Symbol & Data source & Description \\
\midrule
\endhead
\midrule
\multicolumn{4}{r@{}}{\textit{continued on next page}} \\
\endfoot
\bottomrule
\endlastfoot
\multicolumn{4}{@{}l}{\textit{Upper-level: charging station choice}} \\
Dist\_OE      & $d_{nit}^{OE}$        & GPS              & Network distance (km) from origin to the station \\
Detour        & $\Delta t_{nit}$      & GPS              & Detour time (min) required to reach the station \\
Price         & $Price_{i}$           & Charging Station & Charging price (\$/kWh) at the station \\
Num\_L2       & $Num_{i}^{L_{2}}$     & Charging Station & Number of Level 2 charging ports at the station \\
Num\_DCFC     & $Num_{i}^{DCFC}$      & Charging Station & Number of DC fast-charging ports at the station \\
ChargePoint   & $Network_{i}^{Car}$   & Charging Station & Charging network indicator ($=1$ ChargePoint, 0 otherwise) \\
Tesla         & $Network_{i}^{Tes}$   & Charging Station & Charging network indicator ($=1$ Tesla, 0 otherwise) \\
Other         & $Network_{i}^{Oth}$   & Charging Station & Charging network indicator ($=1$ other types, 0 otherwise); ``unknown'' is the reference level \\
Log\_sum      & $IV_{nit}$            & ---              & Logsum utility from the lower-level \\
\addlinespace
\multicolumn{4}{@{}l}{\textit{Lower-level: activity choice}} \\
Fitness       & $POI_{i}^{Fit}$ & SafeGraph POI & Number of fitness POIs within 200 m of the station \\
Pharmacies    & $POI_{i}^{Pha}$ & SafeGraph POI & Number of pharmacy POIs within 200 m of the station \\
Restaurants   & $POI_{i}^{Res}$ & SafeGraph POI & Number of restaurant POIs within 200 m of the station \\
Retail        & $POI_{i}^{Ret}$ & SafeGraph POI & Number of retail POIs within 200 m of the station \\
Grocery       & $POI_{i}^{Gro}$ & SafeGraph POI & Number of grocery POIs within 200 m of the station \\
Beauty        & $POI_{i}^{Bea}$ & SafeGraph POI & Number of beauty-service POIs within 200 m of the station \\
Errands       & $POI_{i}^{Err}$ & SafeGraph POI & Number of errand POIs within 200 m of the station \\
Entertainment & $POI_{i}^{Ent}$ & SafeGraph POI & Number of entertainment POIs within 200 m of the station \\
Beautiful     & $SV_{i}^{Bea}$  & GSV panorama  & Perceived beauty of the built environment surrounding the station \\
Boring        & $SV_{i}^{Bor}$  & GSV panorama  & Perceived boringness of the built environment surrounding the station \\
Depressing    & $SV_{i}^{Dep}$  & GSV panorama  & Perceived depressingness of the built environment surrounding the station \\
Lively        & $SV_{i}^{Liv}$  & GSV panorama  & Perceived liveliness of the built environment surrounding the station \\
Safety        & $SV_{i}^{Saf}$  & GSV panorama  & Perceived safety of the built environment surrounding the station \\
Wealthy       & $SV_{i}^{Wea}$  & GSV panorama  & Perceived wealth of the built environment surrounding the station \\
Stay\_dur     & $\tau_{nit}$    & GPS           & Available charging/activity time after selecting the station for charging \\
\addlinespace
\multicolumn{4}{@{}l}{\textit{$z$ variables: driver and session characteristics}} \\
EV\_share     & $EVS_{n}$ & GPS            & Share (\%) of EV drivers in the user's home CBG \\
INC           & $INC_{n}$ & GPS + Census   & Median household income (\$/year) of the user's home CBG \\
Weekend       & $WKD_{n}$ & GPS            & Weekend indicator for the charging session ($=1$ for weekend, 0 weekday) \\
Daytime       & $DAT_{n}$ & GPS            & Daytime indicator for the charging session ($=1$ if the session occurs between 8:00 a.m. and 8:00 p.m., 0 otherwise) \\
Morning\_peak & $MOP_{n}$ & GPS            & Morning-peak indicator for the charging session ($=1$ if the session occurs between 8:00 and 10:00 a.m., 0 otherwise) \\
OD\_length    & $ODL_{n}$ & GPS            & Network distance (km) from the trip origin to the destination \\
\end{xltabular}
\endgroup

\bibliographystyle{elsarticle-harv}
\bibliography{references}

@misc{fhwa2022,
  author       = {{Federal Highway Administration}},
  title        = {Historic step: All fifty states plus {D.C.} and {Puerto Rico} greenlit to move {EV} charging networks forward, covering 75,000 miles of highway},
  howpublished = {\url{https://highways.dot.gov/newsroom/historic-step-all-fifty-states-plus-dc-and-puerto-rico-greenlit-move-ev-charging-networks}},
  year         = {2022},
  note         = {Accessed 23 September 2026},
}

@misc{whitehouse2023,
  author       = {{The White House}},
  title        = {{FACT SHEET}: {Biden}-{Harris} administration announces new standards and major progress for a made-in-{America} national network of electric vehicle chargers},
  howpublished = {\url{https://bidenwhitehouse.archives.gov/briefing-room/statements-releases/2023/02/15/fact-sheet-biden-harris-administration-announces-new-standards-and-major-progress-for-a-made-in-america-national-network-of-electric-vehicle-chargers/}},
  year         = {2023},
  note         = {Accessed 23 September 2026},
}

@article{pierce2023home,
  title={Home charging access and the implications for charging infrastructure costs in the United States},
  author={Pierce, L and Slowik, P},
  journal={International Council on Clean Transportation},
  year={2023}
}

@article{dot2023charger,
  title={Charger types and speeds},
  author={USDOT, US},
  journal={US Department of Transportation, June. https://www. transportation. gov/rural/ev/toolkit/ev-basics},
  year={2023}
}

@article{gazmeh2026misalignment,
  title={Misalignment between infrastructure and mobility patterns constrains effective access to public EV charging},
  author={Gazmeh, Hossein and Qian, Xinwu and Small, Mario L and Wang, Qi and Guo, Yuntao},
  journal={Sustainable Cities and Society},
  pages={107845},
  year={2026},
  publisher={Elsevier}
}

@article{zheng2024effects,
  title={Effects of electric vehicle charging stations on the economic vitality of local businesses},
  author={Zheng, Yunhan and Keith, David R and Wang, Shenhao and Diao, Mi and Zhao, Jinhua},
  journal={Nature communications},
  volume={15},
  number={1},
  pages={7437},
  year={2024},
  publisher={Nature Publishing Group UK London}
}

@article{wu2025planning,
  title={Planning the electric vehicle transition by integrating spatial information and social networks},
  author={Wu, Jiaman and Salgado, Ariel and Gonz{\'a}lez, Marta C},
  journal={Nature Communications},
  volume={16},
  number={1},
  pages={11220},
  year={2025},
  publisher={Nature Publishing Group UK, London}
}

@article{de2025causal,
  title={Causal spillover effects of electric vehicle charging station placement on local businesses: a staggered adoption study},
  author={De Silva, M Mavin and Clark, Callie and Nakayama, Tadachika and Yabe, Takahiro},
  journal={npj Sustainable Mobility and Transport},
  volume={3},
  number={1},
  pages={62},
  year={2026},
  publisher={Nature Publishing Group UK, London}
}

@article{pan2020location,
  title={A location model for electric vehicle (EV) public charging stations based on drivers’ existing activities},
  author={Pan, Long and Yao, Enjian and Yang, Yang and Zhang, Rui},
  journal={Sustainable cities and society},
  volume={59},
  pages={102192},
  year={2020},
  publisher={Elsevier}
}

@article{wang2025joint,
  title={Joint modelling of electric vehicle charging and daily activity scheduling},
  author={Wang, Senlei and Pougala, Janody and Hillel, Tim},
  journal={Transportation Research Part D: Transport and Environment},
  volume={147},
  pages={104911},
  year={2025},
  publisher={Elsevier}
}

@article{huang2026preferences,
  title={Preferences and Willingness to Pay for Private Electric Vehicle Charger Sharing: A Discrete Choice Analysis in China},
  author={HUANG, Xingjun and Ying, LI and Deqiang, FU and Feng, LIU and Cunrui, MA and Lin, ZHOU},
  journal={Transport Policy},
  pages={104309},
  year={2026},
  publisher={Elsevier}
}

@article{yao2026modeling,
  title={Modeling EV charging behavior: The impact of service experience and user heterogeneity},
  author={Yao, Ming and Lu, Xinchun and Cao, Shuchao},
  journal={Travel Behaviour and Society},
  volume={43},
  pages={101216},
  year={2026},
  publisher={Elsevier}
}

@article{potoglou2023public,
  title={Public charging choices of electric vehicle users: A review and conceptual framework},
  author={Potoglou, Dimitris and Song, Rongqiu and Santos, Georgina},
  journal={Transportation Research Part D: Transport and Environment},
  volume={121},
  pages={103824},
  year={2023},
  publisher={Elsevier}
}

@article{sica2025electric,
  title={Electric vehicle recharging options in urban areas: Discrete choice modeling to estimate user preference},
  author={Sica, Lorenzo and Carboni, Angela and Deflorio, Francesco and Botta, Cristiana},
  journal={Sustainable Cities and Society},
  volume={121},
  pages={106175},
  year={2025},
  publisher={Elsevier}
}

@article{visaria2022user,
  title={User preferences for EV charging, pricing schemes, and charging infrastructure},
  author={Visaria, Anant Atul and Jensen, Anders Fjendbo and Thorhauge, Mikkel and Mabit, Stefan Eriksen},
  journal={Transportation Research Part A: Policy and Practice},
  volume={165},
  pages={120--143},
  year={2022},
  publisher={Elsevier}
}

@article{lee2020exploring,
  title={Exploring electric vehicle charging patterns: Mixed usage of charging infrastructure},
  author={Lee, Jae Hyun and Chakraborty, Debapriya and Hardman, Scott J and Tal, Gil},
  journal={Transportation Research Part D: Transport and Environment},
  volume={79},
  pages={102249},
  year={2020},
  publisher={Elsevier}
}

@article{kim2017heterogeneous,
  title={Heterogeneous hazard model of PEV users charging intervals: Analysis of four year charging transactions data},
  author={Kim, Seheon and Yang, Dujuan and Rasouli, Soora and Timmermans, Harry},
  journal={Transportation Research Part C: Emerging Technologies},
  volume={82},
  pages={248--260},
  year={2017},
  publisher={Elsevier}
}

@article{xu2018planning,
  title={Planning for electric vehicle needs by coupling charging profiles with urban mobility},
  author={Xu, Yanyan and {\c{C}}olak, Serdar and Kara, Emre C and Moura, Scott J and Gonz{\'a}lez, Marta C},
  journal={Nature Energy},
  volume={3},
  number={6},
  pages={484--493},
  year={2018},
  publisher={Nature Publishing Group UK London}
}

@article{fu2025activity,
  title={Activity-travel pattern inference based on multi-source big data},
  author={Fu, Xiao and Zhang, Yi and Ort{\'u}zar, Juan de Dios and L{\"u}, Guonian},
  journal={Transport Reviews},
  volume={45},
  number={1},
  pages={26--48},
  year={2025},
  publisher={Taylor \& Francis}
}

@article{ren2022random,
  title={A random-utility-consistent machine learning method to estimate agents’ joint activity scheduling choice from a ubiquitous data set},
  author={Ren, Xiyuan and Chow, Joseph YJ},
  journal={Transportation Research Part B: Methodological},
  volume={166},
  pages={396--418},
  year={2022},
  publisher={Elsevier}
}

@article{ren2025nonparametric,
  title={Nonparametric mixed logit model with market-level parameters estimated from market share data},
  author={Ren, Xiyuan and Chow, Joseph YJ and Bansal, Prateek},
  journal={Transportation Research Part B: Methodological},
  volume={196},
  pages={103220},
  year={2025},
  publisher={Elsevier}
}

@article{swait2023distribution,
  title={Distribution-free estimation of individual parameter logit (IPL) models using combined evolutionary and optimization algorithms},
  author={Swait, Joffre},
  journal={Journal of choice modelling},
  volume={47},
  pages={100396},
  year={2023},
  publisher={Elsevier}
}

@article{wang2020deep,
  title={Deep neural networks for choice analysis: Architecture design with alternative-specific utility functions},
  author={Wang, Shenhao and Mo, Baichuan and Zhao, Jinhua},
  journal={Transportation Research Part C: Emerging Technologies},
  volume={112},
  pages={234--251},
  year={2020},
  publisher={Elsevier}
}

@article{han2022neural,
  title={A neural-embedded discrete choice model: Learning taste representation with strengthened interpretability},
  author={Han, Yafei and Pereira, Francisco Camara and Ben-Akiva, Moshe and Zegras, Christopher},
  journal={Transportation Research Part B: Methodological},
  volume={163},
  pages={166--186},
  year={2022},
  publisher={Elsevier}
}

@article{cheng2026graph,
  title={Graph neural networks for residential location choice: Connection to classical logit models},
  author={Cheng, Zhanhong and Hu, Lingqian and Bu, Yuheng and Zhou, Yuqi and Wang, Shenhao},
  journal={Transportation Research Part B: Methodological},
  volume={209},
  pages={103464},
  year={2026},
  publisher={Elsevier}
}

@article{li2026decomposing,
  title={Decomposing random taste heterogeneity in discrete choice modeling: A mixture density network-embedded logit model},
  author={Li, Xiaodong and Feng, Tao and Rasouli, Soora and Jia, Peng and Kuang, Haibo},
  journal={Transportation Research Part E: Logistics and Transportation Review},
  volume={212},
  pages={104940},
  year={2026},
  publisher={Elsevier}
}

@article{clark2026using,
  title={Using large scale GPS data to reveal EV driver activity patterns beyond charging sessions},
  author={Clark, Callie and Driscoll, Anne and Ren, Xiyuan and Salah, Salsabil and Gonzalez, Marta C and Chow, Joseph YJ and Yabe, Takahiro},
  journal={arXiv preprint arXiv:2606.02848},
  year={2026}
}

@article{chakraborty2019demand,
  title={Demand drivers for charging infrastructure-charging behavior of plug-in electric vehicle commuters},
  author={Chakraborty, Debapriya and Bunch, David S and Lee, Jae Hyun and Tal, Gil},
  journal={Transportation Research Part D: Transport and Environment},
  volume={76},
  pages={255--272},
  year={2019},
  publisher={Elsevier}
}

@article{li2023uncertainty,
  title={Uncertainty of available range in explaining the charging choice behavior of BEV users},
  author={Li, Hao and Yu, Lu and Chen, Yu and Tu, Huizhao and Zhang, Jun},
  journal={Transportation Research Part A: Policy and Practice},
  volume={170},
  pages={103624},
  year={2023},
  publisher={Elsevier}
}

@article{ge2022charging,
  title={Charging behavior modeling of battery electric vehicle drivers on long-distance trips},
  author={Ge, Yanbo and MacKenzie, Don},
  journal={Transportation Research Part D: Transport and Environment},
  volume={113},
  pages={103490},
  year={2022},
  publisher={Elsevier}
}

@article{mehditabrizi2025route,
  title={En route and home proximity in EV charging accessibility: a spatial equity analysis},
  author={Mehditabrizi, Asal and Tahmasbi, Behnam and Namadi, Saeed Saleh and Cirillo, Cinzia},
  journal={Transportation Research Part D: Transport and Environment},
  volume={146},
  pages={104910},
  year={2025},
  publisher={Elsevier}
}

@article{yang2025dominant,
  title={Dominant charging location choice of commuters and non-commuters: a big data approach},
  author={Yang, Xiong and Zhuge, Chengxiang and Shao, Chunfu and Guo, Runhang and Wong, Andrew Tin Chak and Zhang, Xiaoyu and Sun, Mingdong and Wang, Pinxi and Wang, Shiqi},
  journal={Transportation},
  volume={52},
  number={2},
  pages={439--466},
  year={2025},
  publisher={Springer}
}

@article{wang2026revisiting,
  title={Revisiting electric vehicle charging behavior: Impacts of private shared charging piles on users’ preferences},
  author={Wang, Ke and Liu, Guangchen and Zhao, Jing and Ye, Xin and Pei, Mingyang},
  journal={Travel Behaviour and Society},
  volume={45},
  pages={101359},
  year={2026},
  publisher={Elsevier}
}

@article{gutjar2023configuration,
  title={The configuration of charging stations: What do potential users want?},
  author={Gutjar, Margarita and Kowald, Matthias},
  journal={Travel Behaviour and Society},
  volume={32},
  pages={100579},
  year={2023},
  publisher={Elsevier}
}

@article{gellrich2022spatial,
  title={Spatial and temporal patterns of electric vehicle charging station utilization: a nationwide case study of Switzerland},
  author={Gellrich, Mario and Block, Andreas and Leikert-B{\"o}hm, Ninja},
  journal={Environmental Research: Infrastructure and Sustainability},
  volume={2},
  number={2},
  pages={021003},
  year={2022},
  publisher={IOP Publishing}
}

@article{wang2021electric,
  title={Electric vehicle drivers’ charging behavior analysis considering heterogeneity and satisfaction},
  author={Wang, Yue and Yao, Enjian and Pan, Long},
  journal={Journal of Cleaner Production},
  volume={286},
  pages={124982},
  year={2021},
  publisher={Elsevier}
}

@article{ten2020investigating,
  title={Investigating the future of ultrafast charging: A choice experiment in The Netherlands},
  author={ten Have, Simone Y and Gkiotsalitis, Konstantinos and Geurs, Karst T},
  journal={World Electric Vehicle Journal},
  volume={11},
  number={4},
  pages={70},
  year={2020},
  publisher={MDPI}
}

@article{pan2019modeling,
  title={Modeling EV charging choice considering risk attitudes and attribute non-attendance},
  author={Pan, Long and Yao, Enjian and MacKenzie, Don},
  journal={Transportation Research Part C: Emerging Technologies},
  volume={102},
  pages={60--72},
  year={2019},
  publisher={Elsevier}
}

@article{wen2016modeling,
  title={Modeling the charging choices of battery electric vehicle drivers by using stated preference data},
  author={Wen, Yuan and MacKenzie, Don and Keith, David R},
  journal={Transportation Research Record},
  volume={2572},
  number={1},
  pages={47--55},
  year={2016},
  publisher={SAGE Publications Sage CA: Los Angeles, CA}
}

@article{mcfadden2000mixed,
  title={Mixed MNL models for discrete response},
  author={McFadden, Daniel and Train, Kenneth},
  journal={Journal of applied Econometrics},
  volume={15},
  number={5},
  pages={447--470},
  year={2000},
  publisher={Wiley Online Library}
}

@article{bowman2001activity,
  title={Activity-based disaggregate travel demand model system with activity schedules},
  author={Bowman, John L and Ben-Akiva, Moshe E},
  journal={Transportation research part a: policy and practice},
  volume={35},
  number={1},
  pages={1--28},
  year={2001},
  publisher={Elsevier}
}

@article{train2016mixed,
  title={Mixed logit with a flexible mixing distribution},
  author={Train, Kenneth},
  journal={Journal of choice modelling},
  volume={19},
  pages={40--53},
  year={2016},
  publisher={Elsevier}
}

@article{sifringer2020enhancing,
  title={Enhancing discrete choice models with representation learning},
  author={Sifringer, Brian and Lurkin, Virginie and Alahi, Alexandre},
  journal={Transportation Research Part B: Methodological},
  volume={140},
  pages={236--261},
  year={2020},
  publisher={Elsevier}
}

@misc{cuebiq2026,
  author       = {{Cuebiq}},
  title        = {{Cuebiq} mobility data for research},
  howpublished = {\url{https://www.cuebiq.com/about/data-for-good/}},
  year         = {2025},
  note         = {Accessed 1 March 2026},
}

@misc{afdc2025,
  author       = {{AFDC}},
  title        = {{Alternative Fuels Data Center}: Electric vehicle charging station locations},
  howpublished = {\url{https://afdc.energy.gov/fuels/electricity_locations.html}},
  year         = {2025},
  note         = {Accessed August 2025},
}

@misc{ocm2025,
  author       = {{Open Charge Map}},
  title        = {{Open Charge Map}: Global {EV} charging location data},
  howpublished = {\url{https://www.openchargemap.org}},
  year         = {2025},
  note         = {Licensed under {CC BY} 4.0. Accessed 12 August 2025},
}

@misc{safegraph2022,
  author       = {{SafeGraph}},
  title        = {Global places ({POI}) \& geometry},
  howpublished = {\url{https://doi.org/10.82551/SMXB-1K04}},
  year         = {2022},
  note         = {Distributed by {Dewey Data}. Accessed 12 August 2025},
}

@misc{EVWatts2025,
  author       = {{EV Watts}},
  title        = {Livewire Data Platform},
  howpublished = {\url{https://livewire.energy.gov}},
  year         = {2025},
  note         = {Accessed 1 April 2025},
}

@article{yao2019human,
  title={A human-machine adversarial scoring framework for urban perception assessment using street-view images},
  author={Yao, Yao and Liang, Zhaotang and Yuan, Zehao and Liu, Penghua and Bie, Yongpan and Zhang, Jinbao and Wang, Ruoyu and Wang, Jiale and Guan, Qingfeng},
  journal={International Journal of Geographical Information Science},
  volume={33},
  number={12},
  pages={2363--2384},
  year={2019},
  publisher={Taylor \& Francis}
}

@article{ren2026microtransit,
  title={Microtransit revenue management informed by citywide travel demand and joint subscription--mode choice modeling},
  author={Ren, Xiyuan and Chow, Joseph YJ and Pandey, Venktesh and Yuan, Linfei},
  journal={Transportation Research Part A: Policy and Practice},
  volume={210},
  pages={105046},
  year={2026},
  publisher={Elsevier}
}

@inproceedings{khan2014dbscan,
  title={DBSCAN: Past, present and future},
  author={Khan, Kamran and Rehman, Saif Ur and Aziz, Kamran and Fong, Simon and Sarasvady, Sababady},
  booktitle={The fifth international conference on the applications of digital information and web technologies (ICADIWT 2014)},
  pages={232--238},
  year={2014},
  organization={IEEE}
}

@article{gonzalez2008,
  author  = {Gonz{\'a}lez, Marta C. and Hidalgo, C{\'e}sar A. and Barab{\'a}si, Albert-L{\'a}szl{\'o}},
  title   = {Understanding individual human mobility patterns},
  journal = {Nature},
  year    = {2008},
  volume  = {453},
  number  = {7196},
  pages   = {779--782},
  doi     = {10.1038/nature06958},
}

@book{train2009,
  author    = {Train, Kenneth E.},
  title     = {Discrete Choice Methods with Simulation},
  edition   = {2nd},
  publisher = {Cambridge University Press},
  address   = {Cambridge},
  year      = {2009},
}

@techreport{evwatts2023,
  author      = {{Energetics}},
  title       = {{EV WATTS} Bi-Annual Activity Update: National Charging Station Report},
  institution = {CLEAResult},
  type        = {Project report},
  year        = {2023},
  month       = dec,
  note        = {Electric Vehicle Widescale Analysis for Tomorrow's
                 Transportation Solutions ({EV WATTS}), supported by the
                 U.S. Department of Energy},
  url         = {https://www.clearesult.com/sites/default/files/2024-05/EVWATTS_EVSE_National.pdf},
  urldate     = {2026-11-09}
}

@techreport{usdot2025bca,
  author      = {{USDOT}},
  title       = {Benefit-Cost Analysis Guidance for Discretionary Grant Programs},
  institution = {Office of the Secretary, U.S. Department of Transportation},
  year        = {2024},
  month       = nov,
  note        = {2025 Update},
  url         = {https://www.transportation.gov/sites/dot.gov/files/2024-11/Benefit%20Cost%20Analysis%20Guidance%202025%20Update%20%28Final%29.pdf}
}

@article{xu2017joint,
  title={Joint charging mode and location choice model for battery electric vehicle users},
  author={Xu, Min and Meng, Qiang and Liu, Kai and Yamamoto, Toshiyuki},
  journal={Transportation Research Part B: Methodological},
  volume={103},
  pages={68--86},
  year={2017},
  publisher={Elsevier}
}

@article{zhang2022understanding,
  title={Understanding the generation mechanism of BEV drivers' charging demand: An exploration of the relationship between charging choice and complexity of trip chaining patterns},
  author={Zhang, Yiyuan and Luo, Xia and Qiu, Yuansen and Fu, Yuxue},
  journal={Transportation Research Part A: Policy and Practice},
  volume={158},
  pages={110--126},
  year={2022},
  publisher={Elsevier}
}

@article{ma2022research,
  title={Research on the valley-filling pricing for EV charging considering renewable power generation},
  author={Ma, Shao-Chao and Yi, Bo-Wen and Fan, Ying},
  journal={Energy Economics},
  volume={106},
  pages={105781},
  year={2022},
  publisher={Elsevier}
}

@article{hagerstrand1970people,
  title={What about people in regional science},
  author={H{\"a}gerstrand, Torsten},
  journal={Transport Sociology: Social aspects of transport planning},
  pages={143--158},
  year={1970}
}

@article{miller1991modelling,
  title={Modelling accessibility using space-time prism concepts within geographical information systems},
  author={Miller, Harvey J},
  journal={International Journal of Geographical Information System},
  volume={5},
  number={3},
  pages={287--301},
  year={1991},
  publisher={Taylor \& Francis}
}

@article{kwan1998space,
  title={Space-time and integral measures of individual accessibility: a comparative analysis using a point-based framework},
  author={Kwan, Mei-Po},
  journal={Geographical analysis},
  volume={30},
  number={3},
  pages={191--216},
  year={1998},
  publisher={Wiley Online Library}
}

@article{pendyala2002formulation,
  title={On the formulation of time-space prisms to model constraints on personal activity-travel engagement},
  author={Pendyala, Ram M and Yamamoto, Toshiyuki and Kitamura, Ryuichi},
  journal={Transportation},
  volume={29},
  number={1},
  pages={73--94},
  year={2002},
  publisher={Springer}
}


\end{document}